\documentclass[12pt]{article}
\usepackage{amssymb,amsmath,latexsym}
\usepackage{graphicx}

\begin{document}

\title{The random phase property and the Lyapunov spectrum\\ for disordered multi-channel systems}

\author{Rudolf A. R\"omer$^1$, Hermann Schulz-Baldes$^2$
\\
\\
{\small $^1$Department of Physics and Centre for Scientific Computing,}
\\
{\small University of Warwick, Coventry CV4 7AL, United Kingdom}
\\
{\small $^2$Department Mathematik, Universit\"at Erlangen-N\"urnberg, Germany}
}



\date{ }

\newtheorem{theo}{Theorem}
\newtheorem{defini}{Definition}
\newtheorem{proposi}{Proposition}
\newtheorem{lemma}{Lemma}
\newtheorem{coro}{Corollary}
\newtheorem{rem}{Remark}
\newtheorem{ex}{Example}

\newcommand{\CM}{{\mathbb C}}
\newcommand{\NM}{{\mathbb N}}
\newcommand{\RM}{{\mathbb R}}
\newcommand{\SM}{{\mathbb S}}
\newcommand{\TM}{{\mathbb T}}
\newcommand{\ZM}{{\mathbb Z}}
\newcommand{\LM}{{\mathbb L}}
\newcommand{\KM}{{\mathbb K}}
\newcommand{\HM}{{\mathbb H}}
\newcommand{\DM}{{\mathbb D}}
\newcommand{\GM}{{\mathbb G}}
\newcommand{\UM}{{\mathbb U}}
\newcommand{\WM}{{\mathbb W}}
\newcommand{\FM}{{\mathbb F}}
\newcommand{\IM}{{\mathbb I}}
\newcommand{\Aa}{{\cal A}}
\newcommand{\Pp}{{\cal P}}
\newcommand{\PP}{{\bf P}}
\newcommand{\HH}{{\bf H}}
\newcommand{\UU}{{\bf U}}
\newcommand{\pp}{{\bf p}}
\newcommand{\EE}{{\bf E}}
\newcommand{\Bb}{{\cal B}}
\newcommand{\Dd}{{\cal D}}
\newcommand{\Ff}{{\cal F}}
\newcommand{\Gg}{{\cal G}}
\newcommand{\Ww}{{\cal W}}
\newcommand{\Ss}{{\cal S}}
\newcommand{\Oo}{{\cal O}}
\newcommand{\Tr}{\mbox{\rm Tr}}
\newcommand{\Tt}{{\cal T}}
\newcommand{\Rr}{{\cal R}}
\newcommand{\Nn}{{\cal N}}
\newcommand{\Mm}{{\cal M}}
\newcommand{\Cc}{{\cal C}}
\newcommand{\Jj}{{\cal J}}
\newcommand{\Ii}{{\cal I}}
\newcommand{\Ll}{{\cal L}}
\newcommand{\Qq}{{\cal Q}}
\newcommand{\Kk}{{\cal K}}
\newcommand{\Hh}{{\cal H}}
\newcommand{\cor}{{\mbox{\rm\tiny cor}}}
\newcommand{\av}{{\mbox{\rm\tiny av}}}
\def\essinf{\mathop{\rm ess\,inf}}
\def\esssup{\mathop{\rm ess\,sup}}
\newcommand{\Zh}{{\widehat{Z}}}
\newcommand{\one}{{\bf 1}}
\newcommand{\nul}{{\bf 0}}
\newcommand{\inv}{{\mbox{\rm\tiny inv}}}
\newcommand{\up}{{\mbox{\rm\tiny upp}}}

\maketitle

\begin{abstract}
A random phase property establishing a link between quasi-one-dimensional random Schr\"odinger operators and full random matrix theory is advocated. Briefly summarized it states that the random transfer matrices placed into a normal system of coordinates act on the isotropic frames and lead to a Markov process with a unique invariant measure which is of geometric nature. On the elliptic part of the transfer matrices, this measure is invariant under the full hermitian symplectic group of the universality class under study. While the random phase property can up to now only be proved in special models or in a restricted sense, we provide strong numerical evidence that it holds in the Anderson model of localization. A main outcome of the random phase property is a perturbative calculation of the Lyapunov exponents which shows that the Lyapunov spectrum is equidistant and that the localization lengths  for large systems in the unitary, orthogonal and symplectic ensemble differ by a factor 2 each. In an Anderson-Ando model on a tubular geometry with magnetic field and spin-orbit coupling, the normal system of coordinates is calculated and this is used to derive explicit energy dependent formulas for the Lyapunov spectrum.

\vspace{.1cm}


\end{abstract}



\section{Introduction}

The quantum mechanics of an electron in a disordered wire or a mesoscopic system is described by a random Hamiltonian which, in the tight-binding approximation, is a matrix typically given as the sum of a fixed kinetic part $H_0$ and a random part $\lambda H_1$ containing a coupling constant $\lambda\geq 0$. The location of the random and deterministic matrix entries then reflects the spacial structure of the sample. A widely used paradigmatic model is the Anderson Hamiltonian \cite{Ander}. Calculating analytically physical quantities such as the density of states (DOS), the localization length, or the transmission amplitudes from a given random Hamiltonian is a difficult endeavor. It usually becomes feasible only under the assumption that the randomness of the Hamiltonian somehow leads to one of the classical random matrix ensembles. The latter have invariance properties which allow to calculate the relevant averages. The most bold such assumption states that the Hamiltonian of the sample itself is drawn from a random matrix ensemble, i.e.\ the scattering matrix is given by one of the circular ensembles. This may be justified when studying universal properties both in the metallic phase and the localization phase but  it is not valid in situations where spatial structure is important or where one wants to analyze a dependence on parameters of the model.

\vspace{.1cm}

In this work, we consider quasi-one-dimensional random systems described by random transfer matrices $\Ss=\Ss(\lambda)$ which may or may not be obtained from a random Hamiltonian $H_0+\lambda H_1$. The main aim is then to establish a natural link to invariant ensembles of random matrix theory. In a strictly one-dimensional case, this link degenerates to the random phase approximation \cite{ATAF} which amounts to supposing that the phases of the Dyson-Schmidt variables \cite{Dys,Sch} (called Pr\"ufer phases in the mathematical literature) are distributed according to the Lebesgue measure on the unit circle, which is also the unitary group $\mbox{\rm U}(1)$ of dimension $1$. As is well-known \cite{PF}, this only holds in the regime of weak disorder and if one uses the correct coordinates, namely, the modified Pr\"ufer variables. Our non-commutative generalization to a situation with $L$ channels uses matrix-valued Pr\"ufer phases taking values in a unitary group. These unitaries are in one-to-one correspondence with the isotropic frames (orthonormal coordinate systems within a Lagrangian plane w.r.t.\ the hermitian symplectic form) which hence carry a natural measure. This measure is, moreover, invariant under the natural action of the hermitian symplectic group on the isotropic frames (see Section~\ref{sec-action} for details). Now, the transfer matrices $\Ss(\lambda)$ are in this hermitian symplectic group and certain cocycles associated to the action give the Lyapunov spectrum, a fact at the base of the {\it transfer matrix method} \cite{MK,PS} which is the prime numerical procedure used for the calculation of the Lyapunov spectrum. Our {\it random phase property} (RPP) describes the distribution of the isotropic frames in the weak-coupling limit $\lambda\to 0$ when the transfer matrix is brought into a hermitian symplectic normal form: 

\begin{equation}
\label{eq-normform}
\Mm^{-1}\,\Ss\,\Mm
\;=\;
\Rr\,e^{\lambda \Pp}\;.
\end{equation}
Here $\Mm$ is a symplectic basis change to the normal form $\Rr$ which is a rotation matrix on the elliptic channels and a simple linear expansion on the hyperbolic ones, and $\Pp$ is the random perturbation. Then the RPP states that the invariant measure on the isotropic frames in this coordinate system is given by the above invariant measure on the elliptic channels, combined with a deterministic distribution on the hyperbolic channels. The precise formulation of the RPP is given in Section~\ref{sec-RPP}. Let us note that  one also finds the terms open and closed channels instead of elliptic and hyperbolic channels in the literature \cite{Ben}.

\vspace{.1cm}

There is a different widely used non-commutative generalization of the one-dimen\-sio\-nal random phase approximation going back to the works \cite{Dor,MPK}. It supposes that the unitary matrices in the polar decomposition of mesoscopic blocks of transfer matrices are distributed according to the Haar measure. This is referred to as local maximum entropy Ansatz (MEA) (the global one refers to the transfer matrices of the whole sample) or also isotropy assumption \cite{Ben}. In the weak disorder regime it allows to deduce the DMPK equation for the flow of transmission amplitudes \cite{Dor,MPK}. The latter lead to reasonable physical predictions of universal nature, but without any parameter dependence (such as energy and coupling constants). There are also numerous physics papers discussing the validity of (variants of) the MEA itself (see \cite{MS,MSt,MT,MC,CB,FYMS} and many others), but we are unaware of numerical tests of its validity for concrete models.

\vspace{.1cm}

The basic but crucial difference between our RPP and the local MEA is the following (see Section~\ref{sec-MEA} for details): in the MEA the distribution of the transfer matrices themselves is supposed to be of maximal entropy, whereas in this paper the distribution of the transfer matrices is given and fixed by the model under study and then only the associated random dynamical system on the space of isotropic frames is supposed to have an invariant measure of maximal entropy (that is, a Haar measure). The RPP is considerably weaker than the MEA (see Section~\ref{sec-MEA}), but it nevertheless allows to deduce the Lyapunov spectrum and its energy and model dependence. Of course, we recover the universal features (here the equidistance of the Lyapunov spectrum and the dependence of the localization length on the universality class). Moreover, the RPP provides a concrete statement that is easily falsifiable by numerical analysis. Numerics also allow to estimate the (mesoscopic) length scale beyond which the RPP is valid. We dare to call the RPP a property for the following reasons: The RPP was rigorously proved to hold in the Wegner $L$-orbital model \cite{SS1}. Furthermore, for an Anderson model on a strip, which is precisely the model studied later in the introduction and in Sections~\ref{sec-normal} to \ref{sec-slab} below, rigorous proof could be provided  \cite{SB1} that at least the low moments coincide with those calculated from the RPP (this is sufficient for the calculation of the Lyapunov exponents). Moreover, we will provide numerical evidence that the full RPP does hold for this model.

\vspace{.1cm}

The main application of the RPP presented here concerns the perturbative calculation of the $L$ positive Lyapunov exponents $\gamma_1\geq\ldots\geq\gamma_L\geq 0$. Their definition is recalled in Section~\ref{sec-action}. The localization length is the inverse of the smallest positive Lyapunov exponent $\gamma_L$. In Section~\ref{sec-perturb} we prove a general perturbative formula for the Lyapunov exponents (Theorem~\ref{theo-equidistance}) whenever the RPP holds. It shows that the Lyapunov spectrum is equidistant and that the signature of the universality class (orthogonal, unitary or symplectic) are characteristic factors $2$ between the localization lengths, in agreement with \cite{Dor,EL,MC}. As these universality classes are closely linked to the complex, real and quaternion number field respectively \cite{Meh}, we will consistently index various objects by a number field $\KM$ which is either $\CM$, $\RM$ or $\HM$.

\vspace{.1cm}

Rather than stating Theorem~\ref{theo-equidistance} in this introduction, let us highlight the corollaries for the Anderson-Ando Hamiltonian \cite{Ander,Ando} on a discretized tube. The Hilbert space describing a particle with spin on the tube is $\ell^2 (\ZM,\CM^L)\otimes \CM^2$ where we think of $\CM^L$ as discretized annulus and $\CM^2$ is the spin degree of freedom. On $\ell^2 (\ZM,\CM^L)\otimes \CM^2$ acts the right shift $S_1$ on $\ZM$ and the cyclic shift $S_2$ on the fiber $\CM^L$, namely $(S_2)^L=\one$. Neither of the shift operators $S_1$ and $S_2$ effects the spin degree of freedom. The magnetic spin-orbit Laplacian (in Landau gauge) is a bounded operator on $\ell^2 (\ZM,\CM^L)\otimes \CM^2$ given by
\begin{equation}
\label{eq-kinetic}
H_0
\;=\;
S_1+S_1^*
+e^{\imath\varphi} S_2
+e^{-\imath\varphi}S_2^*
+2\,\imath\,t\,(S_1-S_1^*)\otimes s^y
+
2\,\imath\,t\,(S_2-S_2^*)
\otimes s^x
\end{equation}
Here $\varphi\in [0,2\pi)$ is half the magnetic flux through a cell of the lattice, $t$ is a coupling constant of the spin-orbit coupling and the spin matrices ${\bf s}=(s^x,s^y,s^z)$ are as usual given by $s^x=\frac{1}{2}\left(\begin{smallmatrix} 0 & 1 \\ 1 & 0 \end{smallmatrix}\right)$,
$s^y=\frac{1}{2}\left(\begin{smallmatrix} 0 & -\imath \\ \imath & 0 \end{smallmatrix}\right)$ and $s^z=\frac{1}{2}\left(\begin{smallmatrix} 1 & 0 \\ 0 & -1 \end{smallmatrix}\right)$. Furthermore, the random potential of Anderson type is given by
\begin{equation}
\label{eq-potential}
H_1\;=\;\sum_{n\in\ZM}\,\sum_{l=1}^L\,w_{n,l}\;|n,l\rangle\langle n,l|
\;,
\end{equation}
where the $w_{n,l}$ are independent and identically distributed centered real variables with unit variance. For sake of simplicity, we suppose the support of the distribution of the $w_{n,l}$ to be compact. The Anderson-Ando Hamiltonian is then $H=H_0+\lambda H_1$. It depends on three parameters $\varphi$, $t$ and $\lambda$. For $\varphi\not= 0$, it is in the unitary universality class with no time-reversal symmetry. If $\varphi=0$ and $t=0$, the spin degree of freedom is effectively suppressed and $H$ is in the orthogonal universality class of time-reversal invariant systems with no or even spin. Finally, if $\varphi=0$ and $t\not =0$, the model is in the symplectic universality class of time-reversal systems with odd spin.

\vspace{.1cm}

\begin{figure}
\begin{center}
\includegraphics[width=0.65\textwidth]{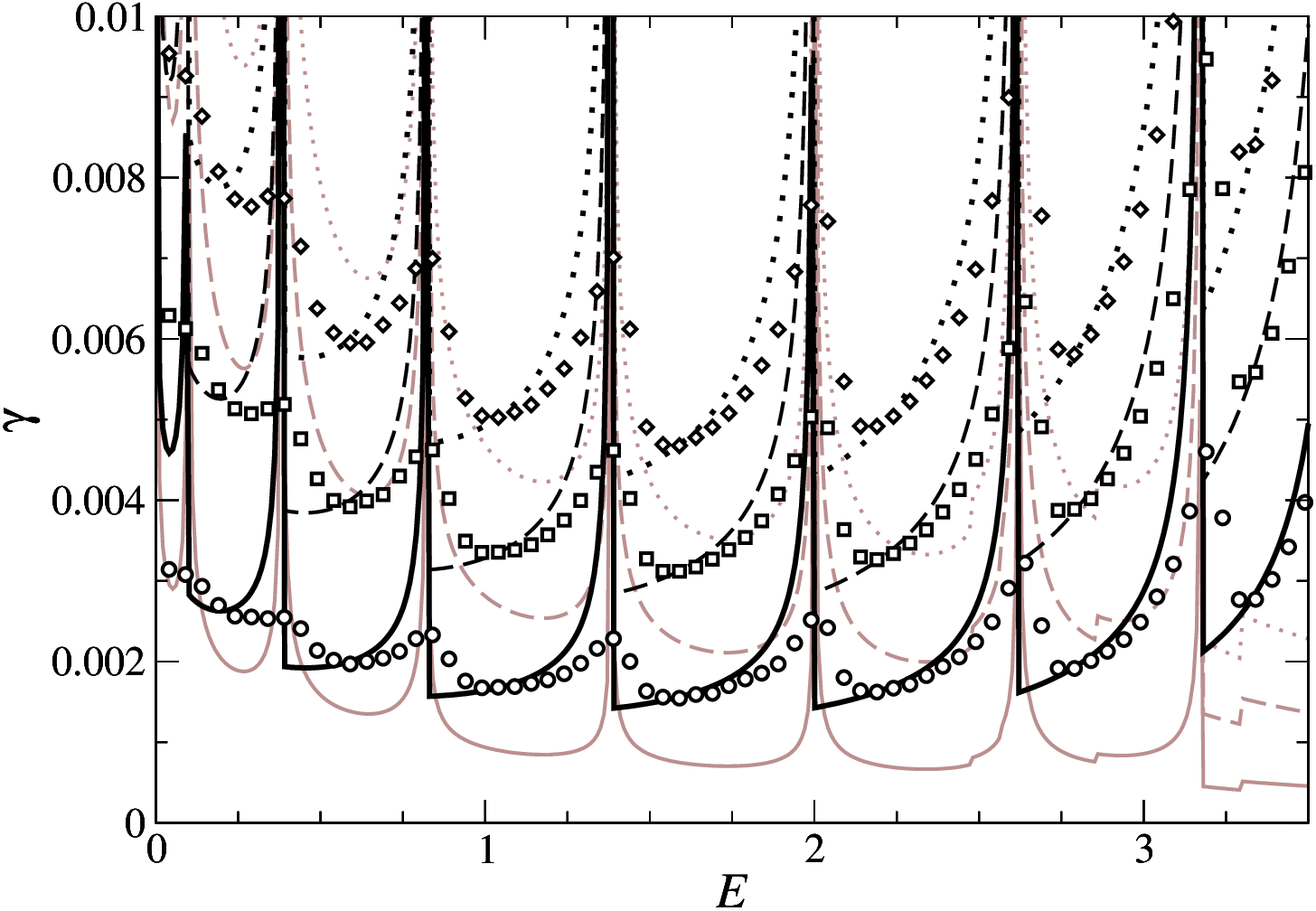}
\caption{\label{fig-1}{
The energy dependence of the smallest three Lyapunov exponents $\gamma_L^\KM$, $\gamma_{L-1}^\KM$ and $\gamma_{L-2}^\KM$ for $\KM=\CM$ {\rm (}grey lines{\rm )} and $\KM=\RM$ {\rm (}black lines{\rm )} as given by {\rm \eqref{eq-gammapcomplex}} and {\rm \eqref{eq-gammapreal}} respectively for a tube width $L=20$ and disorder strength $\lambda=1.11/\sqrt{12}$. For comparison, numerical values calculated by the transfer matrix method (at 1\% error corresponding to one standard deviation) are plotted for $\gamma_{20}^\RM$ ($\circ$), $\gamma_{19}^\RM$ ($\Box$) and $\gamma_{18}^\RM$ ($\diamond$).}}
\end{center}
\end{figure}

As usual \cite{PF,SB2} the (not necessarily square integrable) solutions $\psi$ of the Schr\"o\-dinger equation $H\psi=E\psi$ at energy $E$ can be calculated using transfer matrices. For the Anderson-Ando Hamiltonian and at height $n\in\ZM$, these transfer matrices are
\begin{equation}
\label{eq-transferintro} {\Ss} \;=\; \left(
\begin{array}{cc} [E -(S_2+S_2^*)-t(S_2-S_2^*)2\imath s^x - \lambda \, w_n](1+2t\imath s^y)^{-1} & -(1+2t\imath s^y)^* \\
(1+2t\imath s^y)^{-1} & {0}\end{array} \right) \;,
\end{equation}
where $w_n=\mbox{\rm diag}(w_{n,1},\ldots,w_{n,L})$. This is a $2L\times 2L$ matrix with $2\times 2$ matrix entries describing the spin degree of freedom. In Sections~\ref{sec-normal} to \ref{sec-Ando}, we will construct the hermitian symplectic basis change $\Mm$ needed to bring these transfer matrices into a normal form \eqref{eq-normform}. Both the basis change $\Mm$ and the number $L_e$ of elliptic channels in the normal form $\Rr$ depend on the energy $E$, as do the Lyapunov exponents. Then, supposing that the RPP holds, one can use the general formula given in Theorem~\ref{theo-equidistance} in order to calculate the Lyapunov spectrum. The result is that
\begin{equation}
\label{eq-gammapreal}
\gamma^\RM_p
\;=\;
\frac{\lambda^2}{4L}\;\frac{1}{L_e(L_e+1)}\,
\left[\sum_{l}
\frac{1}{|\sin(k_l)|}\right]^2\;\left(L-p+1\right)\;+\;\Oo(\lambda^3)\;.
\end{equation}
where $k_l=\arccos(\frac{E}{2}-\cos(\frac{2\pi l}{L}))$ for $l=1,\ldots,L$ whenever the argument of the arcus sine is of absolute value less than $1$. The number of $l$'s for which this is the case is precisely the number $L_e$ of elliptic channels and the sum in \eqref{eq-gammapreal} carries precisely over these indices. A similar formula has already been derived by different means in \cite{Dor2}. For the inverse localization length $\gamma^\RM_L$, this formula and its rigorous proof can already be found in our prior works \cite{SB1,RS}, and the case $L=1$ has been known for a long time \cite{PF}. Now for the case without time-reversal invariance, let us take $t=0$ and take the magnetic flux $\varphi$ to be small, but non-vanishing. Then

\begin{equation}
\label{eq-gammapcomplex}
\gamma^\CM_p
\;=\;
\frac{\lambda^2}{4L}\;\frac{1}{L_e^2}\,
\left[\sum_{l}
\frac{1}{|\sin(k_l)|}\right]^2\;\left(L-p+\frac{1}{2}\right)\;+\;\Oo(\varphi \lambda^2,\lambda^3)\;.
\end{equation}
Finally, for the result in the symplectic universality class one takes $\varphi=0$ and $t$ small:
\begin{equation}
\label{eq-gammapquat}
\gamma^\HM_p
\;=\;
\frac{\lambda^2}{4L}\;\frac{1}{L_e(L_e-\frac{1}{2})}\,
\left[\sum_{l}
\frac{1}{|\sin(k_l)|}\right]^2\;\left(L-p+\frac{1}{4}\right)\;+\;\Oo(t\lambda^2,\lambda^3)\;.
\end{equation}
These formulas can be compared to ab initio results obtained with the transfer matrix methods. Figure~\ref{fig-1} shows that the agreement is remarkably good.
Note that each of these formulas have singularities when $k_l$ is close to $0$. This happens at so-called internal band-edges \cite{RS}. The numerics of \cite{RS} and Figure~\ref{fig-1} show that these singularities are washed out. For an analysis of this phenomenon one needs to go beyond the RPP.

\vspace{.1cm}

The first important implication of formulas \eqref{eq-gammapreal}, \eqref{eq-gammapcomplex} and \eqref{eq-gammapquat} is that the elliptic (bottom) part of the Lyapunov spectrum is equidistant with the same spacing. This equidistance also follows from the DMPK equations \cite{MC} and is an important ingredient, for example, the analysis of shot noise \cite{BB}. On a more analytical level, Dorokhov \cite{Dor3} has already shown such equidistance for the Wegner $L$-orbital model (which is closer to random matrix theory because $H_1$ is a full hermitian random matrix), a fact that was rigorously proved in \cite{SS}. The second important conclusion concerns the localization lengths, namely in the limit of large $L_e$, we find

$$
\frac{1}{2}\,\gamma_L^\RM
\;=\;
\gamma_L^\CM
\;=\;
2\,\gamma_L^\HM
\;.
$$
The first equality reflects the suppression of backscattering in presence of magnetic fields. These formulas agree with prior findings by different methods \cite{EL,Dor,MC}.

\vspace{.1cm}

As was already stressed in \cite{RS}, these formulas hold only in a quasi-one-dimensional situation where the sample is much longer than wide.
All formulas have the same scaling predicted by Thouless \cite{Tho}, namely as the sum over $l$ is of order $L_e$, one has roughly $\gamma^\KM_p\sim \frac{\lambda^2}{L}(L-p+1)$. There is basically only a difference of a factor of $2$ between the localization lengths in the three universality classes, whereas it is well-known that the two-dimensional behavior is dramatically different (localization for $\RM$, delocalization for $\HM$ and Landau levels for $\CM$). This indicates that these two-dimensional behaviors cannot be analyzed perturbatively. Finally, let us direct the interested reader to Section~\ref{sec-slab} where formulas for a quasi-one-dimensional Anderson model with a higher-dimensional fibers are derived.

\vspace{.2cm}

Before beginning the technical part of the paper, a short statement about the mathematical rigor may be at place. We definitely hope that Sections~\ref{sec-action} and \ref{sec-perturbformula} as well as the appendices satisfy high standards in this respect, and also the diagonalizations in Section~\ref{sec-normal} to \ref{sec-Ando} are rigorous. Clearly the RPP itself is not proved here, and therefore the proofs of its implications such as Theorem~\ref{theo-equidistance} have only an algebraic value. However, using these algebraic identities the methods of \cite{SB1} allow to prove \eqref{eq-gammapreal} to \eqref{eq-gammapquat} rigorously up to an unknown density matrix (see \cite{RS}, with the RPP this matrix is the identity), albeit away from the internal band edges and with unsatisfactory error estimates. The main short-coming of the RPP (also on the level of theoretical physics) is the lack of a detailed analysis of the internal band edges \cite{RS}, see the discussion in Section~\ref{sec-RPP}. Therefore the error estimates in \eqref{eq-gammapreal} to \eqref{eq-gammapquat} and Theorem~\ref{theo-equidistance} break down at these internal band edges.

\vspace{.3cm}

\noindent {\bf Acknowledgements:} This work was supported by the DFG. H.S.-B.\ thanks the Instituto de Matematicas in Cuernavaca for an extremely pleasant stay during his sabbatical. R.A.R.\ thanks A.\ Stuart and G.\ Golub for discussions.

\section{Action on isotropic frames and flags}
\label{sec-action}

Let us first introduce the symplectic form $\Jj$, the Lorentz form $\Gg$ and the Cayley
transform $\Jj$ as the following $2L\times 2L$ matrices (matrices of this size
are denoted by mathcal symbols in this work), each composed by 4 blocks of size
$L\times L$:
$$
{\cal J}
\;=\;
\left(
\begin{array}{cc}
{ 0} & -{\bf 1} \\
{\bf 1} & { 0}
\end{array}
\right)
\;,
\quad
{\cal G}
\;=\;
\left(
\begin{array}{cc}
{\bf 1} & { 0} \\
{ 0} & -{\bf 1}
\end{array}
\right)
\;,
\quad
\Cc
\;=\;\frac{1}{\sqrt{2}}\;
\left(
\begin{array}{cc} {\bf 1} & -\,\imath\,{\bf 1} \\
{\bf 1} & \imath\,{\bf 1} \end{array}
\right)
\;,
$$
The following identities will be useful:
\begin{equation}
\label{eq-Cayley}
\Cc\, \Jj\,\Cc^*
\;=\;
\frac{1}{\imath}\;\Gg\;,
\qquad
\overline{\Cc}\, \Jj\,\Cc^*
\;=\;
\frac{1}{\imath}\;\Jj\;.
\end{equation}
Next let us recall that the quaternions $\HM$ are the real span of four units $q_0=1$, $q_1$, $q_2$ and $q_3$ which satisfy the algebraic relations $q_1^2=q_2^2=q_3^2=q_1q_2q_3=-q_0$. We will think of these units to be the following complex $2\times 2$ matrices:
$$
q_0\;=\;
\left(
\begin{array}{cc} 1 & 0 \\
0 & 1 \end{array}
\right)
\;,
\quad
q_1\;=\;
\left(
\begin{array}{cc} \imath & 0 \\
0 & -\imath \end{array}
\right)
\;,
\quad
q_2\;=\;
\left(
\begin{array}{cc} 0 & 1 \\
-1 & 0 \end{array}
\right)
\;,
\quad
q_3\;=\;
\left(
\begin{array}{cc} 0 & \imath \\
\imath & 0 \end{array}
\right)
\;.
$$
Hence a quaternion $a\in\HM$ is a $2\times 2$ matrix of the form $a=\sum_{j=0}^3a_j q_j$ with $a_j\in\RM$. Often $q_0$ will be omitted. The quaternion conjugation is defined by $a^*=a_0-\sum_{j=1}^3a_j q_j$ and this coincides with the adjoint of $a$ as $2\times 2$ matrix. A quaternion matrix $A\in\mbox{\rm Mat}(L\times L,\HM)$ is simply a matrix with quaternion entries, hence also a $2L\times 2L$ matrix with complex entries. The adjoint $A^*$ of this complex $2L\times 2L$ matrix is defined as usual. Note that $A^*$ is also given by the transpose of the quaternion conjugate of $A$.

\vspace{.1cm}

Even though simply expressed in terms of $q_2$, it will be convenient to introduce $I=\left(\begin{smallmatrix} 0 & -1 \\ 1 & 0 \end{smallmatrix}\right)$. The reason is that we will allow $I$ to act on arbitrary complex matrices and that it will be used to implement the time reversal symmetry below. Now a given $a\in\mbox{\rm Mat}(2\times 2,\CM)$ is a quaternion if and only if $a=I^*\overline{a}I$ (here the overline denotes the complex conjugation). Further, somehow abusing notation we also write $I=\mbox{\rm diag}(I,\ldots,I)$ with as many entries as needed in a given situation. Hence $I$ always has an even number of rows and columns and satisfies $I^2=-{\bf 1}$. Now given $A\in\mbox{\rm Mat}(2L\times 2L,\CM)$, one has the equivalence of $A\in\mbox{\rm Mat}(L\times L,\HM)$ and $I^*\overline{A}I=A$.

\vspace{.1cm}

Next let us define the hermitian symplectic groups $\mbox{HS}(2L,\KM)$ for $\KM=\RM,\,\CM,\,\HM$:
$$
\mbox{HS}(2L,\KM)
\;=\;
\left\{
\Tt\in\mbox{Mat}({2L}\times 2L,\KM)\,\left|
\,\Tt^*\Jj\Tt\,=\,\Jj\;
\right.
\right\}
\;.
$$
In the case $\KM=\RM$ the hermitian symplectic group $\mbox{HS}(2L,\RM)$ coincides with the symplectic group $\mbox{SP}(2L,\RM)$, but in the other two cases the use of the adjoint instead of transpose makes a difference. In our prior work \cite{SB2} we nevertheless denoted these groups by $\mbox{SP}(2L,\KM)$ (with, moreover, $L$ replaced by $\frac{L}{2}$ in the case $\KM=\HM$). As this leads to notational conflicts with the literature, we now plan to use the notation $\mbox{HS}(2L,\KM)$. Let us note that
$$
\mbox{HS}(2L,\RM)
\;=\;
\left\{
\Tt\in\mbox{HS}({2L},\CM)\,\left|
\,\overline{\Tt}\,=\,\Tt\;
\right.
\right\}
\;,
$$
and similarly $\mbox{HS}(2L,\HM)$ is a subgroup of $\mbox{HS}(4L,\CM)$:
$$
\mbox{HS}(2L,\HM)
\;=\;
\left\{
\Tt\in\mbox{HS}({4L},\CM)\,\left|
\,I^*\overline{\Tt}I\,=\,\Tt\;
\right.
\right\}
\;.
$$
The equations $\overline{\Tt}=\Tt$ and $I^*\overline{\Tt}I=\Tt$ can respectively also be written as $\Tt^t\Jj\Tt=\Jj$ and $\Tt^tI\Jj\Tt=I\Jj$, which were used as definition in \cite{SB2}. Let us also note that $\mbox{HS}(2L,\HM)$ is also isomorphic to the classical group $\mbox{SO}^*(4L)$. For our purposes below, it will at times be convenient to use the Cayley transform of these groups
$$
\mbox{U}(L,L,\KM)\;=\;
\Cc\,\mbox{HS}(2L,\KM)\,\Cc^*
\;.
$$
These groups are called generalized Lorentz groups of signature $(L,L)$, also called pseudo-unitary groups \cite{MPK}. In fact, due to \eqref{eq-Cayley}, $\mbox{U}(L,L,\CM)$ is the group of matrices $\Tt\in\mbox{Mat}({2L},\CM)$ satisfying $\Tt^*\Gg\Tt=\Gg$ and hence conserving the Lorentz form $\Gg$. In $\mbox{U}(L,L,\RM)$ they further satisfy either $\Gg\Jj\overline{\Tt}\Gg\Jj=\Tt$ (or alternatively $\Tt^t\Jj\Tt=\Jj$), while $\mbox{U}(L,L,\HM)$ is the subgroup of $\mbox{U}(2L,2L,\CM)$ characterized by $I^*\Gg\Jj\overline{\Tt}\Gg\Jj I=\Tt$ (or alternatively $\Tt^tI\Jj\Tt=I\Jj$). Unfortunately, $\mbox{U}(L,L,\RM)$ and $\mbox{U}(L,L,\HM)$ are not real and  quaternion matrices other than the notations suggest. This can be considered a disadvantage of the groups $\mbox{U}(L,L,\KM)$, on the other hand the associated frames are simpler than those in the hermitian symplectic representation.

\vspace{.1cm}

One can view $\Jj$ as a sesquilinear (hermitian) form on $\CM^{2L}$. A subspace of $\CM^{2L}$ is called isotropic (w.r.t. $\Jj$) if this sesquilinear form vanishes on it, that is $v^*\Jj w=0$ for all vectors $v,w$ in the subspace. The maximal dimension of an isotropic subspace is $L$ and such maximal isotropic subspaces are also called hermitian Lagrangian. By definition, each element of $\mbox{HS}(2L,\CM)$ maps an isotropic subspace to an isotropic subspace. A flag of isotropic subspaces is an increasing and maximal sequence of isotropic subspaces. The flag manifold $\FM(L,\CM)$ is by definition the set of flags of isotropic subspaces w.r.t. $\Jj$.  Now $\FM(L,\RM)$ is the subset of $\FM(L,\CM)$ composed of flags satisfying $v^t\Jj w=0$ for all vectors $v,w$ in the flag. Similarly $\FM(L,\HM)$ is the subset of $\FM(2L,\CM)$ with flags satisfying $v^tI\Jj w=0$ for all vectors $v,w$. The group $\mbox{HS}(2L,\KM)$ naturally acts on $\FM(L,\KM)$.

\vspace{.1cm}

It will be very convenient to cover $\FM(L,\KM)$ by a simpler object. Each flag in $\FM(L,\KM)$ can be described by a $2L\times L$ matrix $\Phi=(\phi_1,\ldots,\phi_L)$ by letting the $p$th isotropic subspace be spanned by the vectors $\phi_1,\ldots,\phi_p$.
Moreover, one can choose these vectors to be orthonormalized. Hence let us introduce the set of (maximal) isotropic frames w.r.t. $\Jj$ by
\begin{equation}
\label{eq-isoframesHS1}
\IM(L,\KM)\;=\;\left\{\Phi\in\mbox{\rm Mat}(2L\times L,\KM)\,|\,\Phi^*\Phi=\one\;,\Phi^*\Jj\Phi=0\, \right\}\;.
\end{equation}
For sake of concreteness, let us note
$$
\IM(L,\RM)\;=\;\left\{\Phi\in\IM(L,\CM)
\,|\,\overline{\Phi}=\Phi\, \right\}\;,
\qquad
\IM(L,\HM)\;=\;\left\{\Phi\in\IM(2L,\CM)
\,|\,I^*\overline{\Phi}I=\Phi\, \right\}\;.
$$

\vspace{.1cm}

Now, we will also work with the flag manifolds $\Cc \,\FM(L,\KM)$ of isotropic flags w.r.t. the sesquilinear form $\Gg$ satisfying possibly the real or quaternion symmetry.  In the same way, $\Cc\,\IM(L,\KM)$ contains the isotropic frames w.r.t. $\Gg$. These sets are particularly simple. In fact, because $\Cc\,\IM(L,\CM)$ are those $L$-dimensional frames in $\CM^{2L}$ satisfying $\Phi^*\Gg\Phi=0$, one readily verifies that
\begin{equation}
\label{eq-flagrepres}
\Cc\,\IM(L,\CM)\;=\;\left\{\,\left.\frac{1}{\sqrt{2}}\left(
\begin{array}{c}U \\ V
\end{array}
\right)
\,\right|\,U,V\in \mbox{\rm U}(L)\, \right\}\;.
\end{equation}
This also allows to write out a formula for elements of $\IM(L,\CM)$ in terms of two unitaries. Furthermore, the relations $\overline{\Cc^*\Phi}=\Cc^*\Phi$ and
$I^*\overline{\Cc^*\Phi}I=\Cc^*\Phi$ respectively lead to
\begin{equation}
\label{eq-realflagrepres}
\Cc\,\IM(L,\RM)\;=\;\left\{ \,\left.\frac{1}{\sqrt{2}}\left(
\begin{array}{c}U \\ \overline{U}
\end{array}
\right)
\,\right|\,U\in \mbox{\rm U}(L)\,\right\}\;,
\quad
\Cc\,\IM(L,\HM)\;=\;\left\{ \,\left.\frac{1}{\sqrt{2}}\left(
\begin{array}{c}U \\ I^*\overline{U}I
\end{array}
\right)
\,\right|\,U\in \mbox{\rm U}(2L)\,\right\}\;.
\end{equation}
Because of these representations there are natural measures on $\Cc\,\IM(L,\KM)$ (and hence also $\IM(L,\KM)$) induced by the Haar measures on the unitary groups.
Let us note that under the stereographic projection $\Phi=\left(\begin{smallmatrix} a \\ b \end{smallmatrix}\right)\in\IM(L,\KM)\mapsto (a-\imath b)(a+\imath b)^{-1}$ \cite{SB2} these measures give those of Dyson's circular ensembles CUE, COE and CSE respectively on the maximally isotropic subspaces. We will next analyze the projection from $\IM(L,\KM)$ to $\FM(L,\KM)$ and then these measures also lead to natural measures on $\FM(L,\KM)$.

\vspace{.1cm}

Two frames $\Phi,\Phi'\in\IM(L,\CM)$ describe the same flag if and only if $\Phi=\Phi' S$ for un upper triangular matrix $S$. But $\Phi^*\Phi=\one$ and $(\Phi')^*\Phi'=\one$ imply $S^*S=\one$. The only upper triangular unitaries are the diagonal unitaries, which in turn can be identified with the torus $\TM^L$. Hence we conclude that the isotropic frames form a $\TM^L$-cover of the flag manifold:
\begin{equation}
\label{eq-cover}
\FM(L,\CM)\;=\;\IM(L,\CM)\,/\,\TM^{L}\;.
\end{equation}
For $\KM=\RM$, the diagonal unitary $S$ has to be real so that $S\in (\ZM_2)^{L}$, while for $\KM=\HM$ it must satisfy $I^*\overline{S}I={S}$ and the set of such diagonal unitaries is in bijection with the torus $\TM^{L}$. Thus
$$
\FM(L,\RM)\;=\;\IM(L,\RM)\,/\,(\ZM_2)^{L}\;,
\qquad
\FM(L,\HM)\;=\;\IM(L,\HM)\,/\,\TM^{L}\;.
$$

\vspace{.1cm}

Clearly the group $\mbox{\rm HS}(2L,\KM)$ naturally acts on $\FM(L,\KM)$. Besides the simple expressions in \eqref{eq-flagrepres} and \eqref{eq-realflagrepres}, the main reason for introducing $\IM(L,\KM)$ is that the action of $\mbox{\rm HS}(2L,\KM)$ lifts in a natural way which is particularly useful for the calculation of Lyapunov exponents. Let us first study the action of the unitaries in $\mbox{\rm HS}(2L,\KM)$, or equivalently of the unitaries in $\mbox{\rm U}(L,L,\KM)$ on $\Cc\,\IM(L,\KM)$. For $\KM=\CM$, one has $\mbox{\rm U}(L,L,\CM)\cap \mbox{\rm U}(2L)=\mbox{\rm diag} (\mbox{\rm U}(L), \mbox{\rm U}(L))$ and the action on $\Cc\,\IM(L,\CM)$ is simply given by left multiplication. By \eqref{eq-flagrepres} the measure on $\Cc\,\IM(L,\CM)$ induced by the Haar measure on $\mbox{\rm U}(L)\times \mbox{\rm U}(L)$ is invariant under this action. Furthermore, each $\Tt\in\mbox{\rm U}(L,L,\RM)\cap \mbox{\rm U}(2L)\cong\mbox{\rm U}(L)$ is of the form $\Tt=\mbox{\rm diag}(W,\overline{W})$ with $W\in\mbox{\rm U}(L)$ and, similarly, $\Tt\in\mbox{\rm U}(L,L,\HM)\cap \mbox{\rm U}(4L)\cong\mbox{\rm U}(2L)$ is of the form $\Tt=\mbox{\rm diag}(W,I^*\overline{W}I)$ with $W\in\mbox{\rm U}(2L)$ (this can be checked using Lemma~2 of \cite{SB2}). Again, it follows from \eqref{eq-realflagrepres} that the invariant measures on $\Cc\,\IM(L,\RM)$ and $\Cc\,\IM(L,\HM)$ are invariant under these actions. Clearly this transposes to the actions of the unitaries in $\mbox{\rm HS}(2L,\KM)$ acting on $\IM(L,\KM)$.

\vspace{.1cm}

Now we want to extend the action of the unitaries in $\mbox{HS}(2L,\KM)$ to a group action of all of $\mbox{HS}(2L,\KM)$ on $\IM(L,\KM)$.  Given $\Tt\in\mbox{HS}(2L,\KM)$ and $\Phi\in\IM(L,\KM)$, the $2L\times L$ matrix $\Tt\Phi$ is isotropic w.r.t. $\Jj$, but not orthonormalized. However, applying a Gram-Schmidt procedure one then obtains a new isotropic frame $\Tt\cdot\Phi\in \IM(L,\KM)$. In order to write this out more explicitly, let  $\Phi=(\phi_1,\ldots,\phi_L)$ with $\phi_l\in\KM^{2L}$. Then $\Tt\Phi$ is isotropic w.r.t. $\Jj$ and the action is defined by $\Phi'=\Tt\cdot\Phi$ where $\Phi'=(\phi'_1,\ldots,\phi'_L)$ is calculated iteratively by
\begin{equation}
\label{eq-GramSchmidt}
\phi'_p
\;=\;
\frac{\phi''_p}{\|\phi''_p\|}\;,
\qquad
\phi''_p
\;=\;
\Tt\phi_p-\sum_{q=1}^{p-1} \phi'_q\;(\phi'_q)^*\Tt\phi_p
\;.
\end{equation}
Note that $(\phi'_q)^*\Tt\phi_p\in\KM$ and that the order of the factors in the last term is important in the case $\KM=\HM$. Now for $\Tt\in\mbox{U}(L,L,\KM)$ and $\Phi\in\Cc\,\IM(L,\KM)$, we define $\Tt\cdot \Phi=\Cc(\Cc^*\Tt\Cc\cdot \Cc^*\Phi)$.

\vspace{.1cm}

\begin{proposi}
\label{prop-action}
Let $\Tt\in \mbox{\rm HS}(2L,\KM)$ and  $\Phi\in\IM(L,\KM)$.


\noindent {\rm (i)}
There exists a unique upper triangular matrix $S(\Tt,\Phi)\in\mbox{\rm Mat}(L\times L,\KM)$ with positive entries

on the diagonal such that
$$
\Tt\cdot\Phi\;=\;\Tt\,\Phi\,S(\Tt,\Phi)^{-1}\;.
$$

\vspace{.1cm}

\noindent {\rm (ii)} The map $(\Tt,\Phi)\mapsto S(\Tt,\Phi)$ is a multiplicative cocycle for the action of $\mbox{\rm HS}(2L,\KM)$ on $\IM(L,\KM)$,

namely if also $\Tt'\in\mbox{\rm HS}(2L,\KM)$
$$
S(\Tt\Tt',\Phi)\;=\;S(\Tt,\Tt'\cdot\Phi)\,S(\Tt',\Phi)\;.
$$

\vspace{.1cm}

\noindent {\rm (iii)}
This cocycle does not project to a cocycle on $\FM(L,\KM)$, but one has for $T\in\TM^L$
\begin{equation}
\label{eq-cocyclequotient}
S(\Tt,\Phi T)\;=\;T^{-1}\,S(\Tt,\Phi)\,T\;.
\end{equation}
\end{proposi}

\noindent {\bf Proof.} Parts (i) and (ii) are immediate from the considerations above and for \eqref{eq-cocyclequotient}, one has to analyze
\eqref{eq-GramSchmidt} a bit more closely.
\hfill $\Box$

\vspace{.2cm}

Let $e_p$, $p=1,\ldots,L$, be the standard (real) basis of $\KM^L$. Then
$e_p^*S(\Tt,\Phi)e_p>0$ is the $p$th entry of on the diagonal of $S(\Tt,\Phi)$ and one can define the additive real-valued cocycles
$$
g_p(\Tt,\Phi)\;=\;\tau\;\log\bigl(e_p^*S(\Tt,\Phi)e_p\bigr)\;,
$$
where $\tau=1$ in the cases $\KM=\CM,\RM$, and $\tau=\frac{1}{2}\,\Tr_2$ in the case $\KM=\HM$ if the quaternions are identified with $2\times 2$ matrices (hence $\tau$ just extracts the real coefficient of $q_0$).
It follows from \eqref{eq-cocyclequotient} that this is actually a cocycle on the flag manifold $\FM(L,\KM)$. Note that in the case $\KM=\HM$, $e_p^*S(\Tt,\Phi)e_p$ is a quaternion $2\times 2$ matrix, but by Proposition~\ref{prop-action} the diagonal is real and hence each diagonal entry appears at least twice. It can be convenient to extract the diagonal entry by taking half of the trace of the quaternion.

\vspace{.2cm}

Now given a random process $(\Tt_n)_{n\geq 1}$ in HS$(2L,\KM)$, one can define the growth exponents of these cocycles as usual by
\begin{equation}
\label{eq-lyapdef}
\gamma_p^\KM
\;=\;
\lim_{N\to\infty}\,\frac{1}{N}\;g_p(\Tt_N\cdots\Tt_1,\Phi_0)
\;=\;
\lim_{N\to\infty}\,\frac{1}{N}\;\sum_{n=1}^N\; g_p(\Tt_n,\Phi_{n-1})\;,
\end{equation}
where
\begin{equation}
\label{eq-randdym}
\Phi_n
\;=\;
(\Tt_{n}\cdots\Tt_1)\cdot\Phi_0
\;=\;
\Tt_n\cdot\Phi_{n-1}\;,
\qquad
\Phi_0\in\IM(L,\KM)\;,
\end{equation}
and $\Phi_0$ is an initial condition.
These growth exponents are then actually the usual Lyapunov exponents, namely one can check \cite{SB1} that for almost every initial condition $\Phi=(\phi_1,\ldots,\phi_L)$
$$
\sum_{q=1}^p\gamma^\KM_q\;=\;\lim_{N\to\infty}\;\frac{1}{2N}\,\log\,{\det}_p\left(
\langle \phi_l|\Tt_N\cdots\Tt_1|\phi_k\rangle_{1\leq l,k\leq p}\right)\;.
$$
Let us point out again that in the case $\KM=\HM$, \eqref{eq-lyapdef} only defines the $L$ possibly distinct non-negative Lyapunov exponents. The Lyapunov spectrum is indeed twice degenerate and obtained by doubling each of these exponents \cite{SS1}.

\section{The random phase property}
\label{sec-RPP}

In many models of solid state physics one is naturally led to study the random action of hermitian symplectic transfer matrices on flag manifolds and the isotropic frames. Examples are provided in Sections~\ref{sec-normal} to \ref{sec-slab}. This leads to a Markov process on the isotropic frames of the type given in \eqref{eq-randdym}. In a quasi-one-dimensional situation there are associated invariant measures on the isotropic frames and it is typically very difficult to determine them explicitly in an analytic manner. However, in a perturbative situation of weak and isotropic coupling of the randomness we believe that there is a unique such measure which is, moreover, close to a geometric invariant measure, provided one chooses normal coordinates. The random phase property (RPP) makes this claim precise and verifiable. Below we state in which cases and to what extend the RPP can actually be proved and below we provide numerical evidence that the RPP holds in a variety of particular models. In Section~\ref{sec-perturb} we show what kind of conclusions can be drawn from the RPP.

\vspace{.1cm}

We consider the following set-up. Let be given a random process $(\Tt_n)_{n\geq 1}$ in $\mbox{\rm HS}(2L,\KM)$ in the normal form
\begin{equation}
\label{eq-normalform}
\Tt_n\;=\;\Rr\, e^{\lambda \Pp_n}\;,
\end{equation}
where $\lambda\geq 0$ is a coupling constant and $\Pp_n\in \mbox{\rm hs}(2L,\KM)$ are independent and identically distributed Lie algebra elements, and $\Rr\in\mbox{\rm HS}(2L,\KM)$ is a hermitian symplectic matrix of the form $\Rr=\Rr_e\Rr_h=\Rr_h\Rr_e$ where $\Rr_h$ is an expansion matrix and $\Rr_e\in\mbox{\rm O}(2L)$ is a rotation matrix. In order to describe these matrices more explicitly, let $\kappa=\mbox{\rm diag}(\kappa_1,\ldots,\kappa_L)$ with diagonal entries $\kappa_l\in\KM$ ordered according to their modulus and such that there is an $L_h$ with $|\kappa_{L_h}|>1$ and $\kappa_l=1$ for $l={L_h+1},\ldots,L$. Further be given a real matrix $\eta=\mbox{\rm diag}(\eta_1,\ldots,\eta_L)$ such that $\eta_l=0$ for $l=1,\ldots,L_h$. Then the commuting matrices $\Rr_h$ and $\Rr_e$ are given by

$$
\Rr_h\;=\;
\left(
\begin{array}{cc}
\kappa & 0 \\
0 & \frac{1}{\kappa}
\end{array}
\right)
\qquad
\Rr_e\;=\;\left(
\begin{array}{cc}
\cos(\eta)& -\sin(\eta) \\
\sin(\eta) & \cos(\eta)
\end{array}
\right)
\;.
$$
%
%
%
We call $L_h$ the number of hyperbolic channels, and $L_e=L-L_h$ the number of elliptic channels. Then
let $\pi_e$ be $L\times L$ projection matrix of rank $L_e$ such that $\kappa\pi_e=\pi_e$ and let $\pi_h=\one_L-\pi_e$. One has $L_h=\dim(\pi_h)$ and $L_e=\dim(\pi_e)$. We also set $\Pi_h=\mbox{\rm diag}(\pi_h,\pi_h)$ and $\Pi_e=\mbox{\rm diag}(\pi_e,\pi_e)$. They are called the projections on the hyperbolic and elliptic channels respectively. One has $\Rr_e\Pi_h=\Pi_h$ and $\Rr_h\Pi_e=\Pi_e$.
Here the $q$th channel, $q=1,\ldots,L$, is the span of $e_q,e_{L+q}\in\KM^{2L}$ (where $e_q$, $q=1,\ldots,L$ or $q=1,\ldots,2L$, denotes the standard basis of $\KM^L$ or $\KM^{2L}$).
Thus the $q$th channel is elliptic or hyperbolic pending on whether $\pi_e e_q=e_q$ or $\pi_he_q=e_q$.

\vspace{.1cm}

In our examples in Sections~\ref{sec-normal} to \ref{sec-slab} below, quite some algebraic work is needed in order to bring the transfer matrices into the normal form \eqref{eq-normalform}. Furthermore, this normal form cannot be obtained for all energies, but at so-called internal band edges it would have to contain Jordan blocks. The use of the normal form is explained shortly.

\vspace{.1cm}

Now we consider the random action \eqref{eq-randdym} of the process $(\Tt_n)_{n\geq 1}$ on isotropic frames. Then $(\Phi_n)_{n\geq 1}$ is a Markov process, strictly speaking a family of Markov processes indexed by the coupling constant $\lambda$. According to \eqref{eq-flagrepres}, there are unitaries $U_n$ and $V_n$ such that
\begin{equation}
\label{eq-UVdef}
\Cc\,\Phi_n
\;=\;
\frac{1}{\sqrt{2}}
\left(
\begin{array}{c}
U_n \\
V_n
\end{array}
\right)
\;.
\end{equation}
In the cases $\KM=\RM$ and $\KM=\HM$, one has $V_n=\overline{U_n}$ and $V_n=I^*\overline{U_n}I$ respectively. The RPP describes the distribution of the $\Phi_n$ (or equivalently $U_n$ and $V_n$) based on the following heuristics. The hyperbolic rotation $\Rr_h$ dominates the random dynamics and forces the first frame vectors deterministically into the hyperbolic channels. The elliptic rotation also dominates the randomness, but only generates tori in $\IM(L,\KM)$. The random perturbations $\Pp_n$ change the orientation of the axis of these tori in a diffusive manner. The RPP now states that in a perturbative limit first of all the splitting of hyperbolic and elliptic channels holds, second of all the hyperbolic frame vectors are ordered deterministically according the size of the hyperbolic expansion factors and third of all that the random perturbation is maximally efficient in the sense that the distribution on the elliptic channels has maximal entropy and is given by the Haar measure. Thus the RPP states that the invariant measure of the Markov process $\Phi_n$ is unique and of geometric nature. Averages w.r.t. this invariant measure will be denoted $\EE$. In particular, it allows to calculate Birkhoff averages:
$$
\EE\;f(\Phi)
\;=\;
\lim_{N\to\infty}\;
\frac{1}{N}\;\sum_{n=1}^N\;f(\Phi_n)\;,
\qquad
f\in C(\IM(L,\KM))\;.
$$
In order to describe this invariant measure of the RPP, we consider $\pi_e U \pi_e$, $\pi_e U \pi_h$, $\pi_e V \pi_e$ etc. as matrices of size $L_e\times L_e$, $L_e\times L_h$, $L_e\times L_e$, etc.

\vspace{.2cm}

\noindent {\bf Random Phase Property.} {\sl In the limit $\lambda\to 0$ of weak coupling, the invariant measure of the Markov process on $\IM(L,\KM)$ generated by the normal form matrices {\rm \eqref{eq-normalform}} is unique and, with errors of order $\Oo(\lambda^2)$, given by the following properties:


\vspace{.1cm}

\noindent {\rm (R1)} $U=\pi_e U \pi_e+\pi_h U \pi_h$

\vspace{.1cm}

\noindent {\rm (R2)} $\pi_h U \pi_h$ is deterministic and given by $\pi_h$ unless the moduli of $\kappa_l$, $l=1,\ldots,L_h$, are degenerate.

\vspace{.1cm}

\noindent {\rm (R3)} $\pi_e U \pi_e$ is distributed according to the Haar measure on U$(L_e)$ in the cases $\KM=\RM,\CM$, and 

to U$(2L_e)$ in the cases $\KM=\HM$.

\vspace{.1cm}

\noindent {\rm (R4)} In the case $\KM=\CM$, $U$ and $V$ are independent and identically distributed.

}

\vspace{.2cm}

Let us stress that by (R1) $\pi_e U \pi_h$ and $\pi_h U \pi_e$ vanish with errors of order $\Oo(\lambda^2)$. Both (R1) and (R2) can be proved by the techniques of \cite[Proposition~3]{SB1} and \cite[Lemma~7]{SB2}, with error estimates which are not optimal though. Item (R3) is the central piece of the random phase property. It establishes a connection with random matrix theory (here the Dyson circular ensemble). In the perturbative calculation of the Lyapunov exponents in Section~\ref{sec-perturb} below, one actually does not need the full strength of the (R3), but only uses it to evaluate second and fourth moments of the unitaries as given in Appendix A. These identities for the second and fourth moments hold for a much wider class of distributions than those stated in (R3). The point is though that we believe (R3) to hold and give numerical evidence in Section~\ref{sec-normal}.

\vspace{.2cm}

A serious short-coming of the RPP as stated is that it does not cover the (vicinity of) internal band edges \cite{RS}. Such internal band edges correspond to normal forms having Jordan blocks. When considering families of matrices (typically indexed by the energy), these Jordan blocks appear at points where an elliptic channel passes through a parabolic one to become hyperbolic channels (or inversely). When the RPP is applied to the calculation of the Lyapunov spectrum, this leads to singularities (see formulas \eqref{eq-gammapreal} to
\eqref{eq-gammapquat}) which are effectively smoothed out as show the numerics in \cite{RS}. An rigorous analysis of this smoothing in the one-dimensional case is carried out in \cite{SS3}.

\section{Comparison of the RPP with the MEA}
\label{sec-MEA}

The maximal entropy Ansatz (MEA) \cite{Dor,MPK,MS} is a claim about the matrix entries of the following polar decomposition of the transfer matrix of a sample of length $N$ (which in the local MEA is a building block of a longer sample):
\begin{equation}
\label{eq-polardecomp}
\Cc\,\Tt_N\cdots\Tt_1\,\Cc^*
\;=\;
\left(\begin{array}{cc} u_N & 0 \\ 0 & v_N \end{array}\right)
\left(\begin{array}{cc} \sqrt{1+\Lambda_N} & \sqrt{\Lambda_N} \\ \sqrt{\Lambda_N} & \sqrt{1+\Lambda_N} \end{array}\right)
\left(\begin{array}{cc} u'_N & 0 \\ 0 & v'_N\end{array}\right)
\;.
\end{equation}
Here $\Lambda_N\geq 0$ is a diagonal $L\times L$ matrix with non-decreasing diagonal entries and $u_N$, $v_N$, $u'_N$ and $v'_N$ are in $\mbox{\rm U}(L)$. If $\KM=\RM$ one has $v_N=\overline{u_N}$ and $v'_N=\overline{u'_N}$ while for $\KM=\RM$ one has $v_N=I^*\overline{u_N}I$ and $v'_N=I^*\overline{u'_N}I$. The matrix $\Lambda_N$ is unique, but the polar decomposition is not because diagonal unitary factors (in $\TM^L$) may be added (and further unitary factors if $\Lambda_N$ is degenerate). The nicest feature of the polar decomposition is that the scattering matrix of the sample can be directly read off.

\vspace{.1cm}

The MEA now consists in supposing that for $N$ sufficiently large the unitaries
$u_N$, $v_N$, $u'_N$ and $v'_N$ are Haar distributed and independent, apart from the correlation in the cases $\KM=\RM$ and $\KM=\HM$ described above. This corresponds to a system with only elliptic channels (in fact, a detailed treatment of closed channels is not known to us, but could easily be given along the lines of the last section). Hence (R1) and (R2) become irrelevant and we now argue that the MEA implies (R3) and (R4) of the RPP. Indeed, using \eqref{eq-polardecomp},
$$
\frac{1}{\sqrt{2}}
\left(\begin{array}{c} U_N \\ V_N\end{array}\right)
\;=\;
\Cc\Phi_N
\;=\;
\left(\begin{array}{cc} u_N & 0 \\ 0 & v_N\end{array}\right)
\cdot
\frac{1}{\sqrt{2}}
\left(\begin{array}{c} \widetilde{u}_N \\ \widetilde{v}_N\end{array}\right)
\;=\;
\frac{1}{\sqrt{2}}
\left(\begin{array}{c} u_N \widetilde{u}_N \\ v_N\widetilde{v}_N\end{array}\right)
\;,
$$
where
$$
\frac{1}{\sqrt{2}}
\left(\begin{array}{c} \widetilde{u}_N \\ \widetilde{v}_N\end{array}\right)
\;=\;
\left(\begin{array}{cc} \sqrt{1+\Lambda_N} & \sqrt{\Lambda_N} \\ \sqrt{\Lambda_N} & \sqrt{1+\Lambda_N} \end{array}\right)
\left(\begin{array}{cc} u'_N & 0 \\ 0 & v'_N\end{array}\right)
\cdot
\Cc\,\Phi_0
\;.
$$
Thus, if $u_N$ and $v_N$ are Haar distributed (as they are by the MEA), so are $U_N$ and $V_N$ irrespective of what the distribution of $\widetilde{u}_N$ and $\widetilde{v}_N$ is.

\vspace{.1cm}

In conclusion, the MEA implies the RPP. In both approaches one makes an assumption on the invariant measure of a Markov process associated to a random dynamics generated by the transfer matrices. In the RPP the state space of the Markov process is compact and the dynamics relatively simple as described in Section~\ref{sec-action}, on the other hand in the MEA the state space is non-compact (namely $\mbox{\rm HS}(2L,\KM)$) and the dynamics in the polar representation is complicated, that is, it is relative intricate but possible to calculate $(u_{N+1},v_{N+1},\Lambda_{N+1},u'_{N+1},v'_{N+1})$ from $\Tt_{N+1}$ and
$(u_{N},v_{N},\Lambda_{N},u'_{N},v'_{N})$.

\vspace{-.1cm}

\section{Perturbative formula for the additive cocycles}
\label{sec-perturbformula}

Section~\ref{sec-action} shows that the ergodic limits of the cocycles $g_p$ are precisely the Lyapunov exponents. As our main aim in Section~\ref{sec-perturb} will be to develop a controlled perturbation theory for the Lyapunov exponents, the following proposition will prove to be helpful. It is actually quite easy to derive the lowest order terms of the expansion (and this was done in \cite{SB1}). However, the error estimate below is much better than previous ones.

\begin{proposi}
\label{prop-perturbcocycle}
For $\Pp\in\,\mbox{\rm hs}(2L,\CM)$ and  $\Phi\in\IM(L,\CM)$, we introduce the self-adjoint $L\times L$ matrices
$$
P_1(\Phi)
\;=\;\frac{1}{2}\;
\Phi^*(\Pp+\Pp^*)\Phi\;,\qquad
P_2(\Phi)\;=\;\frac{1}{4}\,
\Phi^*(2\Pp^*\Pp\,+\,\Pp^2\,+\,(\Pp^*)^2)\Phi\;.
$$
Then for $\lambda\in\RM$ and with the notation $p_p=\sum_{q=1}^pe_q\,e_q^*$,
%
$$
g_p(e^{\lambda\Pp},\Phi)
\, =\,
\lambda \,\tau\,e_p^*P_1(\Phi)e_p + 
\lambda^2\,\tau\left[e_p^*P_2(\Phi)e_p\! +\! 
\left(e_p^*P_1(\Phi)e_p\right)^2
\!-\! 2\,e_p^*P_1(\Phi)p_pP_1(\Phi)e_p\right]
+\Oo(\lambda^3)\,,
$$
with an error bound depending only on the norm of $\Pp$ {\rm (}in particular, independent of $L${\rm )}.
\end{proposi}

\noindent {\bf Proof.} Let us begin by calculating upper triangular matrices $S_1,S_2$ with real entries on the diagonal such that $S(e^{\lambda\Pp},\Phi)=\one+\lambda S_1+\lambda^2 S_2+\Oo(\lambda^3)$. The definition of $P_1(\Phi)$ and $P_2(\Phi)$ is such that $|e^{\lambda\Pp}\Phi|^2=\one+2\,\lambda\, P_1(\Phi)
+2\,\lambda^2\, P_2(\Phi)+\Oo(\lambda^3)$. Hence the defining property of $S(e^{\lambda\Pp},\Phi)$ leads to the equation
\begin{eqnarray*}
\one & = &
\left(\one+\lambda S_1^*+\lambda^2 S_2^*\right)^{-1}
\left[\one+2\,\lambda\, P_1(\Phi)
+2\,\lambda^2\, P_2(\Phi)\right]
\left(\one+\lambda S_1+\lambda^2 S_2\right)^{-1}
+\Oo(\lambda^3)
\\
& = &
\one\;+\;\lambda\,\left[2\,P_1(\Phi)-S_1-S_1^*\right]
\\
& & +\;\lambda^2\,\left\{2\,P_2(\Phi)+S_1^*S_1+[S_1-2\,P_1(\Phi)]S_1+S_1^*[S_1^*-2\,P_1(\Phi)]-S_2-S_2^*
\right\}
\,+\,\Oo(\lambda^3)\,.
\end{eqnarray*}
Now we use the following basic fact: the unique upper triangular matrix $S$ with real diagonal satisfying $S+S^*=P$ for some self-adjoint $P$ is given by the strict upper triangle of $P$ plus half the diagonal of $P$. Let $\Xi$ denote the corresponding super-operator, namely $S=\Xi[P]$. With this notation,
$$
S_1=\Xi[2\,P_1(\Phi)]\;,
\qquad
S_2=\Xi\{2\,P_2(\Phi)+S_1^*S_1+[S_1-2\,P_1(\Phi)]S_1+S_1^*[S_1^*-2\,P_1(\Phi)]\}\;.
$$
For any triangular $S$ and any analytic function $f$, one has $e_p^*f(S)e_p=f(e_p^*Se_p)$. Applying this to the logarithm and expanding gives
$$
g_p(e^{\lambda\Pp},\Phi)\;=\;\tau\,e_p^*\left\{\lambda \,S_1+\lambda^2\,S_2-\frac{1}{2}\,\lambda^2\,(S_1)^2\right\}e_p
\,+\,\Oo(\lambda^3)\,.
$$
Using $e_p^*\,\Xi[P]e_p=\frac{1}{2}\,e_p^*Pe_p$, some algebra now leads to
%
$$
g_p(e^{\lambda\Pp},\Phi)
=
\lambda\,\tau\,e_p^*P_1(\Phi)e_p + \frac{\lambda^2}{2}\,\tau\left\{ 
\!+\!e_p^*S_1^*S_1e_p \!+\! [e_p^*P_1(\Phi)e_p]^2
\!-\!4\,\Re e[e_p^*P_1(\Phi)S_1e_p]\right\}
,
$$
again with an error of order $\Oo(\lambda^3)$.
Finally, replacing the identities
%
$$
e_p^*S_1^*S_1e_p
\;=\;
4\,e_p^*P_1(\Phi)p_{p-1}P_1(\Phi)e_p + [e_p^*P_1(\Phi)e_p]^2
\;=\;
4\,e_p^*P_1(\Phi)p_{p}P_1(\Phi)e_p -3\, [e_p^*P_1(\Phi)e_p]^2
\;,
$$
and
$$
e_p^*P_1(\Phi)S_1e_p
\;=\;
2\,e_p^*P_1(\Phi)\pi_{p}P_1(\Phi)e_p - [e_p^*P_1(\Phi)e_p]^2
\;=\;
\Re e[e_p^*P_1(\Phi)S_1e_p]
\;,
$$
completes the proof.
\hfill $\Box$

\section{Perturbative formulas for the Lyapunov exponents}
\label{sec-perturb}

As an implication of the RPP, we show in this section how it allows to derive perturbative formulas for the Lyapunov exponents associated to a random process of the form \eqref{eq-normalform}. For sake of simplicity, we also assume a number of further properties. They are not essential, but simplify the algebra. Hence let us suppose that the random perturbation $\Pp$ is centered and that  $|\kappa_1|>|\kappa_2|>\ldots> |\kappa_{L_h}|>1=\kappa_{L_h+1}=\ldots =\kappa_L$. Such strict inequalities hold generically. Then part (R2) asserts that $\pi_hU\pi_h=\pi_h$.

\vspace{.1cm}

The starting point is the formula \eqref{eq-lyapdef} for the Lyapunov exponents, but we suppose that the Birkhoff sum on the r.h.s. of \eqref{eq-lyapdef} can be calculated using the RPP which provides a measure on $\IM(L,\KM)$. Denote the corresponding average combined with that over $\Tt$ by $\EE$, we therefore have $\gamma_p^\KM=\EE\,g_p(\Tt,\Phi)$. Appealing to Proposition~\ref{prop-action}(ii), we deduce
\begin{equation}
\label{eq-gammapexpan}
\gamma_p^\KM\;=\;\EE\;
g_p(\Rr_h,e^{\lambda\Pp}\cdot\Phi)
\;+\;\EE\;
g_p(e^{\lambda\Pp},\Phi)\;,
\end{equation}
where we used that for the unitary $\Rr_e$ one has $g_p(\Rr_e,\Phi')=0$. The first term can be of order $\Oo(1)$, while the second one is always of order $\Oo(\lambda^2)$ due to Proposition~\ref{prop-perturbcocycle} and because $\Pp$ is centered.

\vspace{.1cm}

Let us begin by considering the first contribution in \eqref{eq-gammapexpan}. For this purpose we need to know the upper triangular matrix $S(\Rr_h,\Phi)$ where $\Phi=\frac{1}{2}\left(\begin{smallmatrix} U+ V \\ \imath(U- V) \end{smallmatrix}\right)\in\IM(L,\KM)$ is given in the representation \eqref{eq-flagrepres} and characterized by
$$
\Rr_h\Phi S(\Rr_h,\Phi)^{-1}
\;=\;
\frac{1}{2}
\left(
\begin{array}{c}
\pi_e(U+V)+\kappa \pi_h(U+ V) \\
\imath\,\pi_e(U- V)+\kappa^{-1} \imath\,\pi_h(U- V)
\end{array}
\right)
S(\Rr_h,\Phi)^{-1}\;\in\;\IM(L,\CM)\;.
$$
Now by (R1) one has $\pi_eU=\pi_eU\pi_e\in\mbox{\rm U}(L_e)$, etc., and $\pi_h U=\pi_h$ and $\pi_h V=\pi_h$, the latter also in the case $\KM=\RM,\HM$ because $\pi_h$ is real and commutes with $I$. Thus $S(\Rr_h,\Phi)=\pi_eS(\Rr_h,\Phi)\pi_e+\pi_hS(\Rr_h,\Phi)\pi_h=\pi_e+\pi_h|\kappa|=|\kappa|$ and
\begin{equation}
\label{eq-hypres}
\gamma_p^\KM\;=\;\ln (|\kappa_p|)\,+\,\Oo(\lambda^2)
\;.
\end{equation}
For a hyperbolic channel $p$, this gives the lowest order term, while for an elliptic channel this lowest order contribution vanishes.

\vspace{.1cm}

We now consider the more interesting case of an elliptic channel $p$. It can be argued \cite{SB1} that the first term in \eqref{eq-gammapexpan} does not contribute even to order $\Oo(\lambda^2)$, and therefore we now focus on the second term in \eqref{eq-gammapexpan}. For this let us use the expansion formula in Proposition~\ref{prop-perturbcocycle}. Because $\Pp$ is centered, we have with errors of order $\Oo(\lambda^3)$,
\begin{equation}
\label{eq-gammaexpand}
\gamma_p^\KM=
\frac{\lambda^2}{4}\EE\,\tau\!\left(
e_p^*\Phi^*(2\Pp^*\Pp\!+\!\Pp^2\!+\!(\Pp^*)^2)\Phi e_p
-
\sum_{q=1}^p
(2-\delta_{q,p})\,
e_p^*\Phi^*(\Pp^*\!+\!\Pp)\Phi
e_q\,e_q^*\Phi^*(\Pp^*\!+\!\Pp)\Phi e_p
\right)
.
\end{equation}
Here the expectation $\EE$ is over the random Lie algebra element $\Pp$ and the distribution of the isotropic frames $\Phi$.
The main result of this section is that the Lyapunov spectrum given by this formula is equidistant at small coupling whenever the distribution of the $\Phi$ is given by the RPP.

\begin{theo}
\label{theo-equidistance}
Let the Lyapunov exponents for the $L_e$ elliptic channels be given by {\rm \eqref{eq-gammaexpand}}, that is $p>L_h$. Suppose the {\rm RPP} holds. Then,
with errors of order $\Oo(\lambda^3)$,
%
$$
\gamma_p^\KM
\;=\;
\frac{\lambda^2}{4L_e(L_e+\delta_{\KM,\RM}-\frac{1}{2}\delta_{\KM,\HM})}
\,
\left(L-p+\frac{1}{2}\,\delta_{\KM,\CM}+\delta_{\KM,\RM}+\frac{1}{4}\,\delta_{\KM,\HM}
\right)
\,\EE\,\Tr[\Pi_e(\Pp^*+\Pp)\Pi_e\Pp\Pi_e]
\;,
$$
where $\delta_{a,b}$ is the Kronecker delta equal to $1$ if $a=b$ and equal to $0$ otherwise and the trace $\Tr$ contains a factor $\frac{1}{2}$ in the case $\KM=\HM$ {\rm (}similar as the $\frac{1}{2}$ contained in $\tau${\rm )}.
\end{theo}

According to \eqref{eq-gammaexpand} one has to calculate averages of certain functions on the isotropic frames. Instead of the self-adjoint operators $2\Pp^*\Pp+\Pp^2+(\Pp^*)^2$ and $\Pp^*+\Pp$, let us consider for any self-adjoint $\Aa,\Bb\in\mbox{\rm Mat}(2L\times 2L,\KM)$ and indices $1\leq p,q\leq L$ the quantities
$$
I^\KM_p(\Aa)\;=\;
\EE\,\tau\,e_p^*\Phi^*\Aa\Phi e_p\;,
\qquad
I^\KM_{p,q}(\Bb)=\EE\,\tau\,e_p^*\Phi^*\Bb\Phi
e_q\,e_q^*\Phi^*\Bb\Phi e_p\;,
$$
where the index $\KM$ indicates the symmetry class. Let us also set $\Aa_e=\Pi_e{\Aa}\Pi_e$ and $\Bb_e=\Pi_e{\Bb}\Pi_e$. As $\Jj$ commutes with $\Pi_e$, it follows $\Bb_e\in\mbox{\rm hs}(2L,\KM)$ whenever $\Bb\in\mbox{\rm hs}(2L,\KM)$.

\begin{lemma}
\label{lem-secondmoments} Let $\Aa,\Bb\in\mbox{\rm Mat}(2L\times 2L,\KM)$ be self-adjoint and suppose that $\Bb\in\mbox{\rm hs}(2L,\KM)$ and $\Bb$ is also in the Lie algebra of the special linear group. Let the {\rm RPP} hold and let $L_h< p,q\leq L$ be elliptic. Then, up to errors $\Oo(\lambda^2)$:
$$
\mbox{\rm (i) }\;
I^\KM_p(\Aa) \; = \;
\frac{\Tr[\Aa_e]}{2\,L_e}\;,
\qquad
\mbox{\rm (ii) }\;\sum_{q'=1}^{L_h}\,I^\KM_{p,q'}(\Bb) \; = \;
\frac{\Tr[\Pi_e\,\Bb\,\Pi_p\,\Bb\,\Pi_e]
}{4\,L_e}\;,
$$
$$
\mbox{\rm (iii) }\;
I^\KM_{p,q}(\Bb) \; = \;
\left(\frac{1}{4L_e^2}\;\delta_{\KM,\CM}
\;+\;
\frac{1+\delta_{p,q}}{4L_e(L_e+1)}\;\delta_{\KM,\RM}
\;+\;
\frac{2-\delta_{p,q}}{4L_e(2L_e-1)}\;\delta_{\KM,\HM}
\right)
\Tr[\Bb_e^2]
\;.
$$
\end{lemma}

\noindent {\bf Proof.} (i) First let us write out $\Phi$ in terms of $U$ and $V$
$$
I^\KM_p(\Aa)\;=\;
\frac{1}{4}\;
\EE\,\Tr\left[\left(
\begin{array}{c}
U+V \\
\imath(U-V)
\end{array}
\right)
^*{\Aa} \left(
\begin{array}{c}
U+V \\
\imath(U-V)
\end{array}
\right)
e_pe_p^*\right]\;.
$$
Note that for $\KM=\HM$ the trace contains a factor $\frac{1}{2}$ stemming from the $\frac{1}{2}$ in the definition of $\tau$.
Now because $p$ is elliptic, by (R1) we can replace ${\Aa}$ by $\Aa_e$ with errors of order $\Oo(\lambda^2)$. Thus
\begin{eqnarray*}
4\,I^\KM_p(\Aa)
& = &
\EE\,\Tr\left[
U^* \left(
\begin{array}{c}
1 \\ \imath
\end{array}
\right)^*
\Aa_e
\left(
\begin{array}{c}
1 \\ \imath
\end{array}
\right)
U
e_pe_p^*\right]
+
\EE\,\Tr\left[
V^* \left(
\begin{array}{c}
1 \\ -\imath
\end{array}
\right)^*
\Aa_e
\left(
\begin{array}{c}
1 \\ -\imath
\end{array}
\right)
V
e_pe_p^*\right]
\\
&  & +\;
\EE\,\Tr\left[
U^* \left(
\begin{array}{c}
1 \\ \imath
\end{array}
\right)^*
\Aa_e
\left(
\begin{array}{c}
1 \\ -\imath
\end{array}
\right)
V
e_pe_p^*\right]
+
\EE\,\Tr\left[
V^* \left(
\begin{array}{c}
1 \\ -\imath
\end{array}
\right)^*
\Aa_e
\left(
\begin{array}{c}
1 \\ \imath
\end{array}
\right)
U
e_pe_p^*\right]
\;.
\end{eqnarray*}
The first summand can now directly be calculated by Lemma~\ref{lem-moments} stated in Appendix A, first for the cases $\KM=\CM,\RM$:
%

$$
\EE\,\Tr\left[
U^* \left(
\begin{array}{c}
1 \\ \imath
\end{array}
\right)^*
\!\Aa_e
\left(
\begin{array}{c}
1 \\ \imath
\end{array}
\right)
U
e_pe_p^*\right]
\,=\,
\frac{1}{L_e}
\Tr\left[
\left(
\begin{array}{c}
1 \\ \imath
\end{array}
\right)^*
\!\Aa_e
\left(
\begin{array}{c}
1 \\ \imath
\end{array}
\right)
\right]
\Tr[e_pe_p^*]
\,=\,
\frac{1}{L_e}\;
\Tr[\Aa_e(\one+\imath\Jj)]
\,.
$$
The second summand gives the same result except for the inverse sign. The sum of the first and second term thus give (i). Indeed, the third and fourth summand vanish, for $\KM=\CM$ because of the independence of $U$ and $V$, and for $\KM=\RM$ because either the factor $U$ appears twice or the factor $\overline{U}$ does, so that again the average over the unitary group vanishes.

\vspace{.1cm}

In the case $\KM=\HM$, the unitary group has dimension $2L_e$, but $\Tr(e_pe_p^*)=2$, so the end result is the same because $\Tr(I^*\Aa I)=\Tr(\Aa)$ for a quaternion matrix $\Aa$.

\vspace{.1cm}

(ii) First let us note that $\sum_{q'=1}^{L_h}e_{q'}e_{q'}^*=\pi_h$. Thus by (R1) and (R2), $\Phi\pi_h\Phi^*=\frac{1}{2}\Pi_h(\one+\Jj\Gg)$, it follows immediately from (i) that (ii)$\,=\frac{1}{4L_e}\Tr(\Pi_e\Bb\Pi_h(\one+\Jj\Gg)\Bb\Pi_e)$.
But $\Bb=\Bb^*\in\mbox{\rm hs}(2L,\KM)$ implies $\Jj\Bb=\Bb\Jj^*$. Therefore $\Jj\Gg\Jj^*=-\Gg$ combined with the cyclicity of the trace implies
$\Tr(\Pi_e\Bb\Pi_h\Jj\Gg\Bb\Pi_e)=-\Tr(\Pi_e\Bb\Pi_h\Jj\Gg\Bb\Pi_e)=0$ so that the result follows.

\vspace{.1cm}

(iii) We now need to calculate fourth moments. Let us first expand in terms of $U$ and $V$ such that $I^\KM_{p,q}(\Bb)$ can be written as
$$
\frac{1}{16}
\EE\Tr\left[\left(
\begin{array}{c}
U+V \\
\imath(U-V)
\end{array}
\right)^*\!{\Bb}_e \left(
\begin{array}{c}
U+V \\
\imath(U-V)
\end{array}
\right)
e_pe_p^*
\left(
\begin{array}{c}
U+V \\
\imath(U-V)
\end{array}
\right)^*\!{\Bb}_e \left(
\begin{array}{c}
U+V \\
\imath(U-V)
\end{array}
\right)
e_qe_q^*\right].
$$
Factoring $U$ and $V$ in each of the four factors, one has $16$ terms. However, all but $6$ of them vanish because one needs the arguments to come in complex conjugate pairs $(U,\overline{U})$ and $(V,\overline{V})$ in order to have a non-vanishing average over the unitary group ({\it cf.} Appendix A). Even though very lengthy, let us write out the result:
\begin{eqnarray}
16\,I^\KM_{p,q}(\Bb) & = &
\EE\,\Tr\left[U^*\left(
\begin{array}{c}
1 \\
\imath
\end{array}
\right)^*\!{\Bb}_e \left(
\begin{array}{c}
1 \\
\imath
\end{array}
\right)
Ue_pe_p^*U^*
\left(
\begin{array}{c}
1 \\
\imath
\end{array}
\right)^*\!{\Bb}_e \left(
\begin{array}{c}
1 \\
\imath
\end{array}
\right)U
e_qe_q^*\right]
\label{eq-number1}
\\
&  & +\;
\EE\,\Tr\left[V^*\left(
\begin{array}{c}
1 \\
-\imath
\end{array}
\right)^*\!{\Bb}_e \left(
\begin{array}{c}
1 \\
-\imath
\end{array}
\right)
Ve_pe_p^*V^*
\left(
\begin{array}{c}
1 \\
-\imath
\end{array}
\right)^*\!{\Bb}_e \left(
\begin{array}{c}
1 \\
-\imath
\end{array}
\right)V
e_qe_q^*\right]
\label{eq-number4}
\\
&  & +\;
\EE\,\Tr\left[U^*\left(
\begin{array}{c}
1 \\
\imath
\end{array}
\right)^*\!{\Bb}_e \left(
\begin{array}{c}
1 \\
\imath
\end{array}
\right)
Ue_pe_p^*V^*
\left(
\begin{array}{c}
1 \\
-\imath
\end{array}
\right)^*\!{\Bb}_e \left(
\begin{array}{c}
1 \\
-\imath
\end{array}
\right)V
e_qe_q^*\right]
\label{eq-number2}
\\
&  & +\;
\EE\,\Tr\left[V^*\left(
\begin{array}{c}
1 \\
-\imath
\end{array}
\right)^*\!{\Bb}_e \left(
\begin{array}{c}
1 \\
-\imath
\end{array}
\right)
Ve_pe_p^*U^*
\left(
\begin{array}{c}
1 \\
\imath
\end{array}
\right)^*\!{\Bb}_e \left(
\begin{array}{c}
1 \\
\imath
\end{array}
\right)U
e_qe_q^*\right]
\label{eq-number3}
\\
&  & +\;
\EE\,\Tr\left[U^*\left(
\begin{array}{c}
1 \\
\imath
\end{array}
\right)^*\!{\Bb}_e \left(
\begin{array}{c}
1 \\
-\imath
\end{array}
\right)
Ve_pe_p^*V^*
\left(
\begin{array}{c}
1 \\
-\imath
\end{array}
\right)^*\!{\Bb}_e \left(
\begin{array}{c}
1 \\
\imath
\end{array}
\right)U
e_qe_q^*\right]
\label{eq-number5}
\\
&  & +\;
\EE\,\Tr\left[V^*\left(
\begin{array}{c}
1 \\
-\imath
\end{array}
\right)^*\!{\Bb}_e \left(
\begin{array}{c}
1 \\
\imath
\end{array}
\right)
Ue_pe_p^*U^*
\left(
\begin{array}{c}
1 \\
\imath
\end{array}
\right)^*\!{\Bb}_e \left(
\begin{array}{c}
1 \\
-\imath
\end{array}
\right)V
e_qe_q^*\right]
\label{eq-number6}
\end{eqnarray}
Now each of these terms can be calculated using  Lemma~\ref{lem-moments}. The matrices $B$ and $D$ in Lemma~\ref{lem-moments} are always given by $B=B^t=e_pe_p^*$ and $D=D^t=e_qe_q^*$, while $A$ and $C$ differ in each of the terms. If we assume that $p\not = q$, one has $BD=BD^t=0$ so that many terms vanish. Furthermore $\Tr(B)=\Tr(D)=1$ if $\KM=\CM,\RM$ and $\Tr(B)=\Tr(D)=2$ if $\KM=\HM$. We will show that \eqref{eq-number1} to \eqref{eq-number3} vanish, and that $\eqref{eq-number5}=\eqref{eq-number6}$ give the contribution on the r.h.s.\ in all cases.

\vspace{.1cm}

We now first focus on the case $\KM=\CM$ and $p\not = q$. We deal with \eqref{eq-number1} using Lemma~\ref{lem-moments}(iii):
$$
\eqref{eq-number1}
\;=\;
\frac{1}{L_e^2-1}\Tr
\left[\left(
\begin{array}{c}
1 \\
\imath
\end{array}
\right)^*\!{\Bb}_e \left(
\begin{array}{c}
1 \\
\imath
\end{array}
\right)
\;
\left(
\begin{array}{c}
1 \\
\imath
\end{array}
\right)^*\!{\Bb}_e \left(
\begin{array}{c}
1 \\
\imath
\end{array}
\right)
\right]
\,-\,
\frac{1}{L_e(L_e^2-1)}\Tr
\left[\left(
\begin{array}{c}
1 \\
\imath
\end{array}
\right)^*\!{\Bb}_e \left(
\begin{array}{c}
1 \\
\imath
\end{array}
\right)
\right]^2
\;.
$$
Hence
$$
\eqref{eq-number1}
\;=\;
\frac{1}{L_e^2-1}\Tr
\left[\Bb_e^2-\Bb_e\Jj\Bb_e\Jj
\right]
\,-\,
\frac{1}{L_e(L_e^2-1)}
\Tr\left[{\Bb}_e(\one+\imath \Jj)
\right]
\;.
$$
However, the last term vanishes because $\Tr(\Bb_e)=\Tr(\Jj\Bb_e)=0$, and $\Jj\Bb_e\Jj=\Bb_e\Jj^*\Jj=\Bb_e$ implies $\Tr[\Bb_e\Jj\Bb_e\Jj]=\Tr[\Bb_e^2]$ so that also the first summand vanishes.
As $V$ has the same distribution as $U$, \eqref{eq-number4} gives the same result as \eqref{eq-number1} except for the minus sign in front of the imaginary units and it thus follows  $\eqref{eq-number4}=0$.
Furthermore, \eqref{eq-number2} is calculated by iterating Lemma~\ref{lem-moments}(i) twice. Thus
$$
\eqref{eq-number2}
\;=\;
\frac{1}{L_e}
\,\Tr
\left[\left(
\begin{array}{c}
1 \\
\imath
\end{array}
\right)^*\!{\Bb}_e \left(
\begin{array}{c}
1 \\
\imath
\end{array}
\right)
\right]
\;
\EE\,\Tr
\left[V^*\left(
\begin{array}{c}
1 \\
-\imath
\end{array}
\right)^*\!{\Bb}_e \left(
\begin{array}{c}
1 \\
-\imath
\end{array}
\right)V
e_qe_q^*e_pe_p^*
\right]
\;.
$$
Thus for $p\not = q$, one immediately sees \eqref{eq-number2}$=0$, but actually this always holds because the first factor vanishes for the same reason as above. Similarly,  \eqref{eq-number3}$=0$. Finally, again using Lemma~\ref{lem-moments}(i) twice,
$$
\eqref{eq-number5}
\;=\;
\frac{1}{L_e^2}
\,
\Tr\left[
\left(
\begin{array}{c}
1 \\
-\imath
\end{array}
\right)^*\!{\Bb}_e \left(
\begin{array}{c}
1 \\
\imath
\end{array}
\right)
\left(
\begin{array}{c}
1 \\
\imath
\end{array}
\right)^*\!{\Bb}_e \left(
\begin{array}{c}
1 \\
-\imath
\end{array}
\right)
\right]
\;=\;
\frac{\Tr[\Bb_e^2+\Bb_e\Jj\Bb_e\Jj]}{L_e^2}
\;=\;
\frac{2\Tr
[\Bb_e^2]}{L_e^2}
\;.
$$
As $\eqref{eq-number5}=\eqref{eq-number6}$, adding up all terms leads to the prefactor for $\KM=\CM$ and $p\not = q$.

\vspace{.1cm}

For $p=q$ each of the six terms above has two supplementary contributions by Lemma~\ref{lem-moments}. Let us start with:
$$
\eqref{eq-number1}
\;=\;
\frac{1}{L_e(L_e+1)}
\left(
\Tr
\left[\left(
\begin{array}{c}
1 \\
\imath
\end{array}
\right)^*\!{\Bb}_e \left(
\begin{array}{c}
1 \\
\imath
\end{array}
\right)
\right]^2
\,+\,
\Tr
\left[\left(
\begin{array}{c}
1 \\
\imath
\end{array}
\right)^*\!{\Bb}_e \left(
\begin{array}{c}
1 \\
\imath
\end{array}
\right)
\;
\left(
\begin{array}{c}
1 \\
\imath
\end{array}
\right)^*\!{\Bb}_e \left(
\begin{array}{c}
1 \\
\imath
\end{array}
\right)
\right]
\;\right)
\;.
$$
As each of the two traces vanishes by the calculation above, $\eqref{eq-number1}=0$ and similarly $\eqref{eq-number4}=0$. We already showed $\eqref{eq-number2}=\eqref{eq-number3}=0$. Finally, \eqref{eq-number5} and \eqref{eq-number6} are the same as in case $p\not = q$ and thus (iii) for $\KM=\CM$ is shown also for $p=q$.

\vspace{.1cm}

Now $\KM=\RM$, which allows us to use for the real and self-adjoint $\Bb$ the identity
\begin{equation}
\label{eq-transposeid}
\left[\,\left(
\begin{array}{c}
1 \\
-\imath
\end{array}
\right)^*\!{\Bb}_e \left(
\begin{array}{c}
1 \\
\imath
\end{array}
\right)\right]^t
\;=\;
\left(
\begin{array}{c}
1 \\
-\imath
\end{array}
\right)^*\!{\Bb}_e \left(
\begin{array}{c}
1 \\
\imath
\end{array}
\right)
\;,
\end{equation}
The evaluation of \eqref{eq-number1} and \eqref{eq-number4} is the same as in the case $\KM=\CM$ and thus $\eqref{eq-number1}=\eqref{eq-number4}=0$. Furthermore Lemma~\ref{lem-moments}(iv) implies that $\eqref{eq-number2}=\eqref{eq-number3}=0$ because $BD=BD^t=0$ for $p\not=q$, while for $p=q$
$$
\eqref{eq-number2}
\;=\;
\eqref{eq-number3}
\;=\;
\frac{1}{L_e(L_e+1)}
\left(
\Tr
\left[\Bb_e^2-\Bb_e\Jj\Bb_e\Jj
\right]
\,+\,
\Tr\left[{\Bb}_e
\right]^2
\,+\,\Tr\left[\Jj{\Bb}_e
\right]^2
\right)\;,
$$
which vanishes again.
For \eqref{eq-number5} and \eqref{eq-number6} we will use Lemma~\ref{lem-moments}(v):
$$
\eqref{eq-number5}
\;=\;
\eqref{eq-number6}
\;=\;
\frac{1}{L_e(L_e+1)}\;
\Tr
\left[\left(
\begin{array}{c}
1 \\
-\imath
\end{array}
\right)^*\!{\Bb}_e \left(
\begin{array}{c}
1 \\
\imath
\end{array}
\right)
\;
\left(
\begin{array}{c}
1 \\
\imath
\end{array}
\right)^*\!{\Bb}_e \left(
\begin{array}{c}
1 \\
-\imath
\end{array}
\right)
\right]
\;=\;
\frac{2\,\Tr[\Bb_e^2]}{L_e(L_e+1)}
\;,
$$
which gives the case $\KM=\RM$ for $p\not =q$. For $p=q$, there are two more contributions in
Lemma~\ref{lem-moments}(v) showing that both $\eqref{eq-number5}$ and $\eqref{eq-number5}$ are exactly twice as large, showing the formula also for $p=q$.

\vspace{.1cm}

In the case $\KM=\HM$, the changes w.r.t. the case $\KM=\RM$ are that now $V=I^*\overline{U}I$ and $\Tr(e_pe_p^*)=2$, and that $2L_e$ replaces $L_e$ as the dimensions on the unitary group. Moreover, we have at our disposal that $I^*\overline{\Bb}I=I^*{\Bb}^tI=\Bb$ for the self-adjoint quaternion matrix $\Bb$ so that \eqref{eq-transposeid} holds with factors of $I$ and $I^*$ from left and right on either one of the identities. The verification of $\eqref{eq-number1}
=\eqref{eq-number4}=0$ is the same as in the case $\RM$, that of
$\eqref{eq-number2}=\eqref{eq-number3}=0$ also modulo the use of the modification of identity \eqref{eq-transposeid}. In the evaluation of \eqref{eq-number5} and \eqref{eq-number6}, all four terms of Lemma~\ref{lem-moments}(v) contribute. With some care one gets the corresponding prefactor.
\hfill $\Box$

\vspace{.1cm}

\noindent {\bf Remark.} As a check of the above formulas it is worth noting that the sum rule
\begin{equation}
\label{eq-sumrule}
\sum_{q'=L_h+1}^{L}\,I^\KM_{p,q'}(\Bb) \; = \;
\frac{\Tr[\Bb_e^2]
}{4\,L_e}\;,
\end{equation}
can be checked independently starting from the identity
$$
\sum_{q'=L_h+1}^L\Phi e_{q'}e_{q'}^*\Phi^*
\;=\;
\Phi\,\pi_e\,\Phi^*
\;=\;
\frac{1}{2}\,\Pi_e\,+\,\frac{1}{2}\,\Cc^*\,
\begin{pmatrix}
0 & U\pi_e V^* \\
V\pi_e U^* & 0
\end{pmatrix}
\,\Cc
\;.
$$
The contribution corresponding to $\frac{1}{2}\Pi_e$ is evaluated using Lemma~\ref{lem-secondmoments}(i) and gives the r.h.s. of \eqref{eq-sumrule}, because the contribution of the second summand vanishes (as shows some further algebra). Then one has, in all three cases of Lemma~\ref{lem-secondmoments}(iii) indeed (as it should be by Lemma~\ref{lem-secondmoments})
$$
\frac{\Tr[\Bb_e^2]}{4\,L_e}
\;=\;
(L_e-1)\,I^\KM_{p,q}\;+\;I^\KM_{p,p}\;,
\qquad
p\not = q\;.
$$

\vspace{.2cm}

\noindent {\bf Proof of Theorem~\ref{theo-equidistance}.}
We need to evaluate each term in \eqref{eq-gammaexpand}. The first term is calculated using
Lemma~\ref{lem-secondmoments}(i) and the identity $\Pp^*=\Jj\Pp\Jj$:
\begin{equation}
\label{eq-intermed}
I_p^\KM[2\Pp^*\Pp+\Pp^2+(\Pp^*)^2]
\;=\;
\frac{1}{L_e}\,
\Tr[\Pi_e(\Pp+\Pp^*)\Pi_e\Pp\Pi_e]
\;+\;\frac{1}{L_e}\,
\Tr[\Pi_e(\Pp+\Pp^*)\Pi_h\Pp\Pi_e]
\;.
\end{equation}
The sum over $q$ is split into a hyperbolic part $1\leq q\leq L_h$ and an elliptic one $L_h+1\leq q\leq L$. The hyperbolic one is given by
$$
\sum_{q=1}^{L_h}\,2\,I^\KM_{p,q}(\Pp^*+\Pp)
\;=\;
\frac{1}{L_e}
\Tr[\Pi_e(\Pp+\Pp^*)\Pi_h\Pp\Pi_e]
\;,
$$
where the equality follows from Lemma~\ref{lem-secondmoments}(ii) and again
$\Pp^*=\Jj\Pp\Jj$. Due to the sign in \eqref{eq-gammaexpand}, this hyperbolic part thus cancels with one of the terms on the r.h.s. of \eqref{eq-intermed}. The elliptic part is given by
$$
\sum_{q=L_h+1}^{p}\,(2-\delta_{p,q})\,I^\KM_{p,q}(\Pp^*+\Pp)
\;=\;
2(p-L_h-1)\,I^\KM_{p,p'}(\Pp^*+\Pp)
-
2\,I^\KM_{p,p}(\Pp^*+\Pp)
\;,
$$
where $p'\not =p$.
Now only remains to evaluate the contributions using Lemma~\ref{lem-secondmoments}(iii) and then carefully resemble all terms. Let us exemplify with the case $\KM=\RM$.
$$
I^\RM_{p,p'}(\Pp^*+\Pp)
\;=\;
2\,
\frac{1+\delta_{p,p'}}{4L_e(L_e+1)}
\;\Tr[\Pi_e(\Pp+\Pp^*)\Pi_e\Pp\Pi_e]
\;,
$$
so that
%
$$
\gamma^\RM_p
\;=\;
\frac{\lambda^2}{4}
\;\EE\,\Tr[\Pi_e(\Pp+\Pp^*)\Pi_h\Pp\Pi_e]
\;
\left[
\frac{1}{L_e}-\frac{2\cdot 2}{4L_e(L_e+1)}\,(p-L_h-1)-\frac{1}{2}\,\frac{2\cdot 2\cdot 2}{4L_e(L_e+1)}
\right]
\;,
$$
which leads to the formula in the case $\KM=\RM$.
\hfill $\Box$

\section{Magnetic Anderson model for tube geometry}
\label{sec-normal}

In order to calculate the Lyapunov spectrum more explicitly than in Theorem~\ref{theo-equidistance}, one needs to include further model specific information. From this section on, we will study the one-particle tight-binding Hamiltonians $H=H_0+\lambda H_1$ described in equations \eqref{eq-kinetic} and \eqref{eq-potential} in the introduction. In this section, we take $\varphi\not = 0$ and $t=0$. Hence the spin degree of freedom is suppressed and can thus be neglected. Therefore the transfer matrices \eqref{eq-transferintro} at energy $E\in\RM$ reduces to an $2L\times 2L$ matrix
\begin{equation}
\label{eq-transfer} {\Ss} \;=\; \left(
\begin{array}{cc} E\,\one -(e^{\imath\varphi} S_2
+e^{-\imath\varphi}S_2^*
+ \lambda \, w_n) & {\bf -1} \\
{\bf 1} & {0}\end{array} \right) \;,
\end{equation}
where the dependence on $n$ is suppressed. The transfer matrix ${\Ss}$ is in the hermitian symplectic group HS$(2L,\CM)$. We first bring the transfer matrix into a normal form. Let us introduce, for $l=1,\ldots,L$,
$$
f_l=\left(\begin{array}{c} f_{l,1} \\ \vdots \\ f_{l,L}
\end{array}\right)
\in\mathbb{C}^L \mbox{ , } \qquad f_{l,k}\;=\;\frac{1}{\sqrt{L}}\;
\exp\left(\frac{2\pi\imath\,lk}{L}\right) \mbox{ . }
$$

\begin{figure}[tb]
\begin{center}
\includegraphics[width=0.65\textwidth]{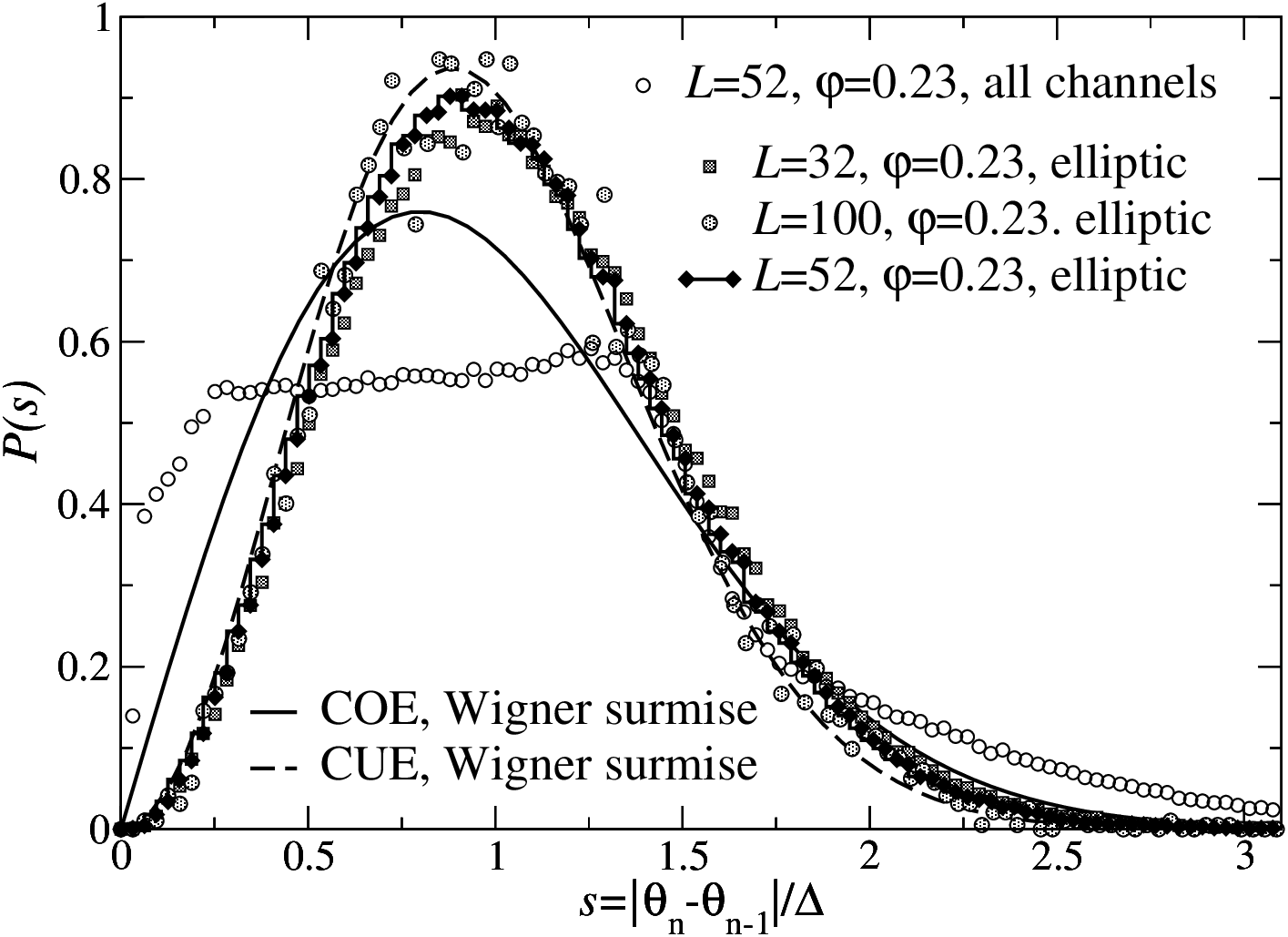}
\caption{\label{fig-2}{
Plot of the level spacing distribution $P(s)$ of the unitary matrices $U_N$ (open circles) and $\pi_e U_N\pi_e$ (black circles) at $\varphi/2\pi=0.23$, $\lambda=0.12/\sqrt{12}$ and $E=1.31$ on slabs of width $L=52$ and length $N=1000$. Data for $\pi_e U_N\pi_e$ with width $L=32$ highlight the increasing agreement with the CUE distribution upon increasing the system size. Data for $L=100$ shows $P(s)$ at $N=1000$ only. Here and in all following figures, $100$ different samples have been averaged over. Also, similar results have been obtained for $L=20$, $32$ and $42$ and $\varphi/2\pi=0.1$, $0.21$, $0.25$, $0.31$, $0.41$, $0.51$, $0.61$, $0.71$, $0.81$ and $0.91$.
}}
\end{center}
\end{figure}

\begin{figure}[tbh]
\begin{center}
\includegraphics[width=0.65\textwidth]{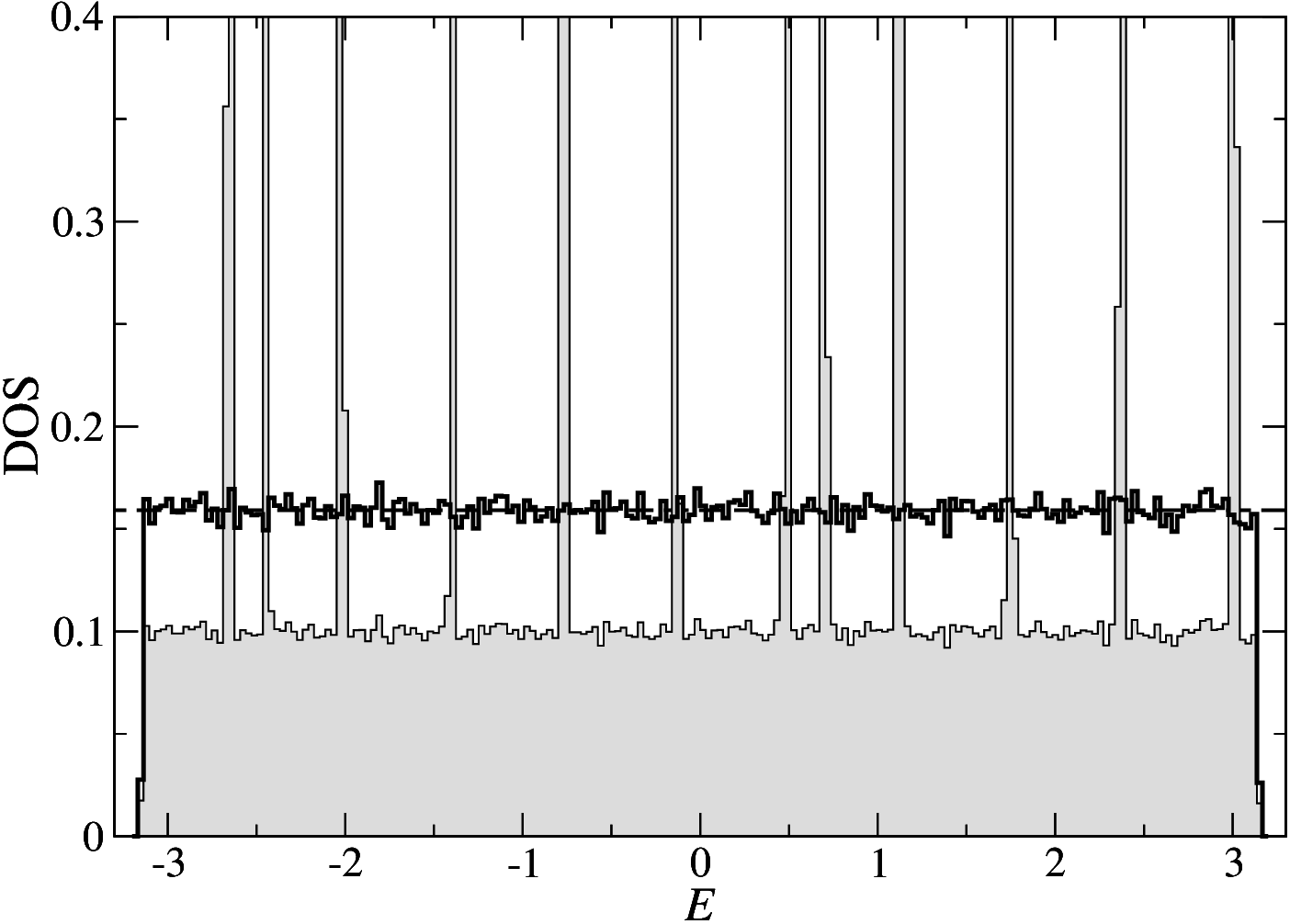}
\caption{\label{fig-3}{
Plot of the distribution of the eigenvalues of the unitary matrices $U_N$ (filled grey histogram) and $\pi_e U_N\pi_e$ (solid black line) at $\varphi/2\pi=0.23$, $\lambda=0.12/\sqrt{12}$ and $E=1.31$ on slabs of width $L=52$ and length $N=1000$. The dashed horizontal line at $1/2\pi$ indicates a normalized, uniform distribution from $-\pi$ to $\pi$.
}}
\end{center}
\end{figure}

\noindent Then we set $m=(f_1,\ldots,f_L)$.
Hence $m$ is the discrete Fourier transform (in U$(L)$) and $\Mm=\mbox{\rm diag}(m,m)\in \mbox{HS}(2L,\CM)\cap\mbox{U}(2L)$. One has $m^*S_2m=e^{\imath\eta}$ with $\eta=\frac{2\pi}{L}\mbox{\rm diag}(1$, $2$, $\ldots, L)$. Therefore
$m^*(E-
e^{\imath\varphi} S_2
-e^{-\imath\varphi}S_2^*)m=E-2\cos(\varphi+\eta)$ is diagonal as well and is non-degenerate (for $\varphi\not =0$). Let $q$ be an adequate $L\times L$ permutation matrix such that the diagonal entries are ordered according to their modulus. Then set

$$
\mu
\;=\;
q^*m^*(E-
e^{\imath\varphi} S_2
-e^{-\imath\varphi}S_2^*)mq
\;,
$$
which hence has its diagonal entries ordered according to their modulus.
Setting $\Qq=\mbox{\rm diag}(q,q)$ it follows that
\begin{equation}
\label{eq-transfer2}
\Qq^{-1}{\cal M}^{-1}{\Ss}{\cal M}\Qq
\;=\; \left(
\begin{array}{cc}  \mu & -{\bf 1} \\
{\bf 1} & 0\end{array} \right) \;\left(
\begin{array}{cc}  {\bf 1} & 0 \\
\lambda\,q^*{W}q & {\bf 1}\end{array} \right)
\;,
\end{equation}
where ${W}=m^*w_nm$, which is a self-adjoint Toeplitz matrix with  first row $(\hat{w}_0, \ldots , \hat{w}_{L-1})$ where $\hat{w}_l=\frac{1}{L}\,\sum_{k=1}^L\,w_{n,k}\,
\exp\left(\frac{2\pi\imath\,lk}{L}\right)$ is the Fourier transform of the potential at height $n$. The energy-dependent $L\times L$ matrix $\mu$ is
diagonal and real for real $E$. Next we want to bring the r.h.s. of \eqref{eq-transfer2}
for $\lambda=0$ into a normal form. The characteristic equation is $\rho^2-\mu\rho+\one=0$ for a diagonal complex matrix $\rho=$diag$(\rho_1,\ldots,\rho_L)$. We choose the first branch of the root and set

$$
\rho\;=\;\frac{\mu}{2}\,+\,\frac{1}{2}\,\sqrt{\mu^2-4}
\;.
$$
If $|\mu_l|<2$, then $|\rho_l|=1$ and $\Im m(\rho_l)>0$ and the $l$th channel is elliptic. If on the other hand $|\mu_l|>2$, then $|\rho_l|>1$ and the channel is hyperbolic. Hence the projections on the hyperbolic and elliptic channels are given by $\pi_h= \chi(|\mu|>2)$ and $\pi_e= \chi(|\mu|<2)$ (here $\chi$ denotes the usual characteristic function of an event). The matrices $\kappa$ and $\eta$ of Section~\ref{sec-RPP} are $\kappa=\rho\pi_h+\pi_e$ and $\eta=\rho\pi_e+\pi_h$.
If $|\mu_l|=1$, then the eigenvalue is degenerate and $\kappa_l=1$ or $\kappa_l=-1$ and the channel is called parabolic; as there is only $1$ eigenvector, the parabolic channel cannot be diagonalized and leads to a Jordan block. We exclude energies which have a parabolic block. As already indicated, such energies are called internal band edges \cite{RS}.

\begin{figure}[tbh]
\begin{center}
\includegraphics[width=0.65\textwidth]{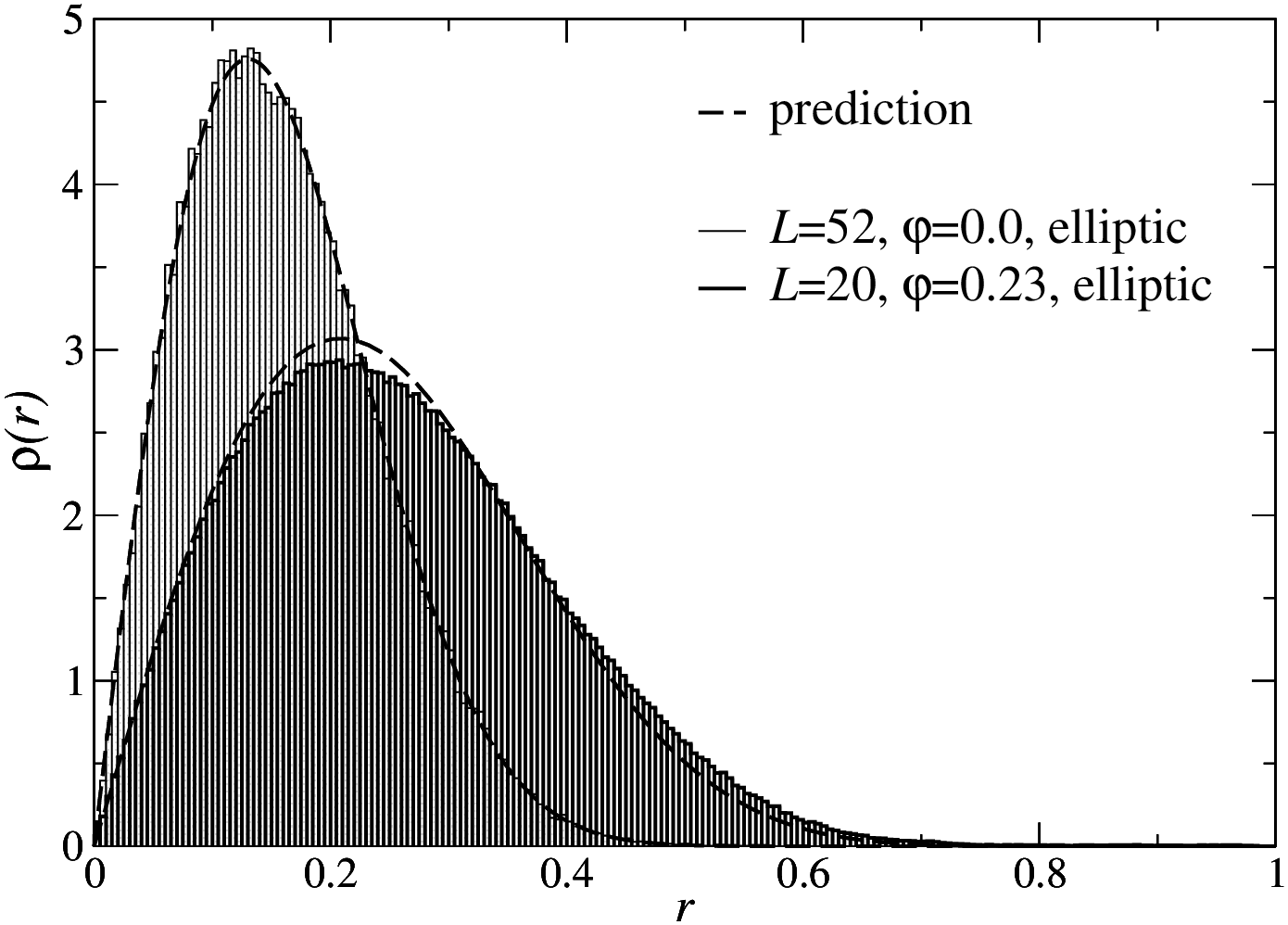}
\caption{\label{fig-4}{
Plot of the distribution $\rho(r)$ of the modulus of the matrix entries of $\pi_e U_N\pi_e$ for the same parameter values as in {\rm Figure~\ref{fig-2}}. For comparison the curve {\rm \eqref{eq-raddistribution}} is plotted as dashed line. Note how the agreement between the numerical data and {\rm \eqref{eq-raddistribution}} gets better upon increasing the system size from $L=20$ ($L_e=13$, front histogram) to $L=52$ ($L_e=31$, back).}}
\end{center}
\end{figure}

\vspace{.1cm}

Next, we introduce the real diagonal matrix $h(\varphi)=((\pi_h +\frac{1}{2\imath} \pi_e)(\rho-\frac{1}{\rho}))^{-\frac{1}{2}}$ depending on the magnetic flux $\varphi$ and then set

$$
\Nn\;=\;
\left(
\begin{array}{cc}
h(\varphi) & h(\varphi)\pi_h \\
h(\varphi)\frac{1}{2}(\rho+\rho^{-1})\pi_e+h(\varphi)\rho^{-1}\pi_h &
h(\varphi)\frac{1}{2\imath}(\rho-\rho^{-1})\pi_e+h(\varphi)\rho\pi_h
\end{array} \right) \;.
$$
One verifies that this matrix is in  HS$(2L,\mathbb{C})$. As $\rho$, $h(\varphi)$, $\pi_e$ and $\pi_h$ all commute, the inverse ${\cal N}^{-1}={\cal J}^*{\cal N}^*{\cal J}$ is thus equal to

$$
\Nn^{-1}\;=\;
\left(
\begin{array}{cc}
h(\varphi)\frac{1}{2\imath}(\rho-\rho^{-1})\pi_e+h(\varphi)\rho\pi_h
& -h(\varphi)\pi_h \\
-h(\varphi)\frac{1}{2}(\rho+\rho^{-1})\pi_e-h(\varphi)\rho^{-1}\pi_h
& h(\varphi)
\end{array} \right) \;.
$$
Now
$$
{\cal N}^{-1}\left(
\begin{array}{cc}  \mu & {\bf -1} \\
{\bf 1} & {0}\end{array} \right){\cal N}
\;=\;
\left(
\begin{array}{cc}  {\rho}\pi_h+\frac{1}{2}(\rho+\rho^{-1})\pi_e
& -\frac{1}{2\imath}(\rho-\rho^{-1})\pi_e \\
\frac{1}{2\imath}(\rho-\rho^{-1})\pi_e &
\rho^{-1} \pi_h+\frac{1}{2}(\rho+\rho^{-1})\pi_e
\end{array} \right)\;.
$$
This matrix is the desired normal form $\Rr=\Rr_h\Rr_e$ with $\kappa$ and $\eta$ as defined above. Now we come back to \eqref{eq-transfer2} and set

$$
\Tt\;=\;
\Nn^{-1}\Qq^{-1}{\cal M}^{-1}{\Ss}\,{\cal M}\Qq\Nn
\;.
$$
Then $\Tt\in\mbox{HS}(2L,\CM)$ and
\begin{equation}
\label{eq-transfer3}
\Tt\;=\;\Rr\, e^{\lambda \Pp}\;,\qquad \mbox{with }\;\;
\Pp\;=\;\Nn^{-1}\left(
\begin{array}{cc} 0 & 0 \\
q^*{W}q & 0\end{array} \right)\Nn\;.
\end{equation}
As $\Nn$ depends on $E$, so does $\Pp$.
Note that $\Pp\in\mbox{hs}(2L,\CM)$ and that $\Pp$ is nilpotent and the only random matrix. This is the desired normal form \eqref{eq-normform} and \eqref{eq-normalform}. Now it makes sense to ask whether the RPP holds for the magnetic Anderson model. We do not have an analytical proof for the RPP, but do provide the following numerical evidence. From \eqref{eq-transfer3} a realization of the Markov process $\Phi_n$ defined in \eqref{eq-randdym} can be generated numerically (with a fixed initial condition $\Phi_0$), giving thus a sequence of unitaries $(U_n,V_n)_{1\leq n\leq N}$ by \eqref{eq-UVdef}. We choose $N=1000$ and thus obtain an empirical ensemble of unitaries $(U_N,V_N)$ consisting of $9000$ matrices from $100$ independent disorder configurations for each of the $90$ intermediate matrices produced during the course of the Markov process after the first $100$ multiplications. As a test whether this ensemble is distributed as stated in the RPP, we plot the level spacing statistics $P(s)$ for this ensemble (see \cite{Meh} for a definition). Figure~\ref{fig-2} shows the result for $U_N$ and $\pi_e U_N\pi_e$ for parameter values as stated in the figure caption. The curves obtained for $V_N$ and $\pi_e V_N\pi_e$ are similar. Hence the results support the central property (R3) of the RPP, as does Figure~\ref{fig-3}. Another numerical check of (R3) is whether the matrix entries of $\pi_e U_N\pi_e$ are distributed on the unit disc $\{re^{\imath\theta}\in\CM\,|\,r\in [0,1],\;\theta\in[0,2\pi)\}$ according to the rotation invariant measure
\begin{equation}
\label{eq-raddistribution}
\frac{L_e-1}{\pi}\;(1-r^2)^{L_e-2}\,r\,dr\,d\theta\;.
\end{equation}
The distribution of the modulus $r$ of the matrix entries of $\pi_e U_N\pi_e$ is shown in Figure~\ref{fig-4}.
Note, however, that clearly the entries are not independent. We also checked that the Hilbert-Schmidt norm of $\pi_e U_N\pi_h$ is very small with high probability which verifies (R1). In conclusion, we consider this as sufficient evidence that the RPP holds at least approximately for the magnetic Anderson model.

\vspace{.1cm}

Now we want deduce formula \eqref{eq-gammapcomplex} for the Lyapunov spectrum from the RPP and Theorem~\ref{theo-equidistance}.
Let us begin by writing out more explicit formulas for $\Pp\in\mbox{hs}(2L,\CM)$, using the notation $W_\varphi=h(\varphi)q^*Wqh(\varphi)$,

$$
\Pp\;=\;
\left(
\begin{array}{cc} -\pi_h W_\varphi & \pi_h W_\varphi\pi_h \\
W_\varphi & W_\varphi\pi_h\end{array} \right)
\;.
$$
From this it follows that
$$
\Pi_e\,\Pp\,\Pi_e\;=\;
\left(
\begin{array}{cc} 0 & 0 \\
\pi_eW_\varphi\pi_e & 0 \end{array} \right)
\;,
$$
so that $(\Pp_e)^2=0$ and $W_\varphi^*=W_\varphi$ imply
$$
\Pi_e(\Pp^*+\Pp)\,\Pi_e\,\Pp\,\Pi_e\;=\;
\left(
\begin{array}{cc} \pi_eW_\varphi^*\pi_e W_\varphi\pi_e & 0 \\
0 & 0 \end{array} \right)
\;.
$$
For the calculation of the Lyapunov spectrum using case $\KM=\CM$ of Theorem~\ref{theo-equidistance}, we now need to calculate the average over $\Pp$.
For this purpose we use
\begin{equation}
\label{eq-Fourierav}
\EE ( \hat{w}_p\hat{w}_q)\;=\;\frac{1}{L}\,\delta_{p+q\,\mbox{\rm\tiny mod}\,L}\;.
\end{equation}
This implies for any diagonal matrix $d=(d_1,\ldots,d_L)$,
$$
\EE\;{W}^*\,d\,{W}
\;=\;
\left(\frac{1}{L}\,\sum_{l=1}^L\,d_l\right)\;\one\;.
$$
Thus
\begin{equation}
\label{eq-expval}
\EE\,\Tr[\Pi_e(\Pp^*+\Pp)\,\Pi_e\,\Pp\,\Pi_e]
\;=\;
\EE\,\Tr( \pi_eW_\varphi\pi_eW_\varphi\pi_e)
\;=\;
\frac{1}{L}\,\left(\sum_{k=L_h+1}^L\,(h_k(\varphi))^2\right)^2
\;.
\end{equation}
It thus follows from Theorem~\ref{theo-equidistance} that, if the RPP holds, for any elliptic index $L_h<p\leq L$ one has
\begin{equation}
\label{eq-gammap}
\gamma^\CM_p
\;=\;
\frac{\lambda^2}{4L}\;\frac{1}{L_e^2}\,
\left[\sum_{k=L_h+1}^L
h_k(\varphi)^2\right]^2\;\left(L-p+\frac{1}{2}\right)+\Oo(\lambda^3)\;.
\end{equation}
Now replacing the definition of $h(\varphi)$ and expanding in $\varphi$ leads to the formula \eqref{eq-gammapcomplex} given in the introduction.

\newpage

\section{Anderson model on tube geometry}
\label{sec-zeromag}

This section deals with the modifications to the prior analysis in the case of vanishing magnetic field $\varphi=0$. The transfer matrix given in \eqref{eq-transfer} now has real entries and $S_2+S_2^*$ has a twofold degenerate spectrum (apart from the top and possibly the bottom eigenvalues). Hence the transfer matrices are in HS$(2L,\RM)$ corresponding to the symmetry class of time-reversal invariant systems with even spin. All the formulas in Section~\ref{sec-normal} remain valid, but a further basis change is needed in order to assure that the new normal form is in HS$(2L,\RM)$. All modified objects will carry a sombrero.

\vspace{.1cm}

\begin{figure}[tb]
\begin{center}
\includegraphics[width=0.65\textwidth]{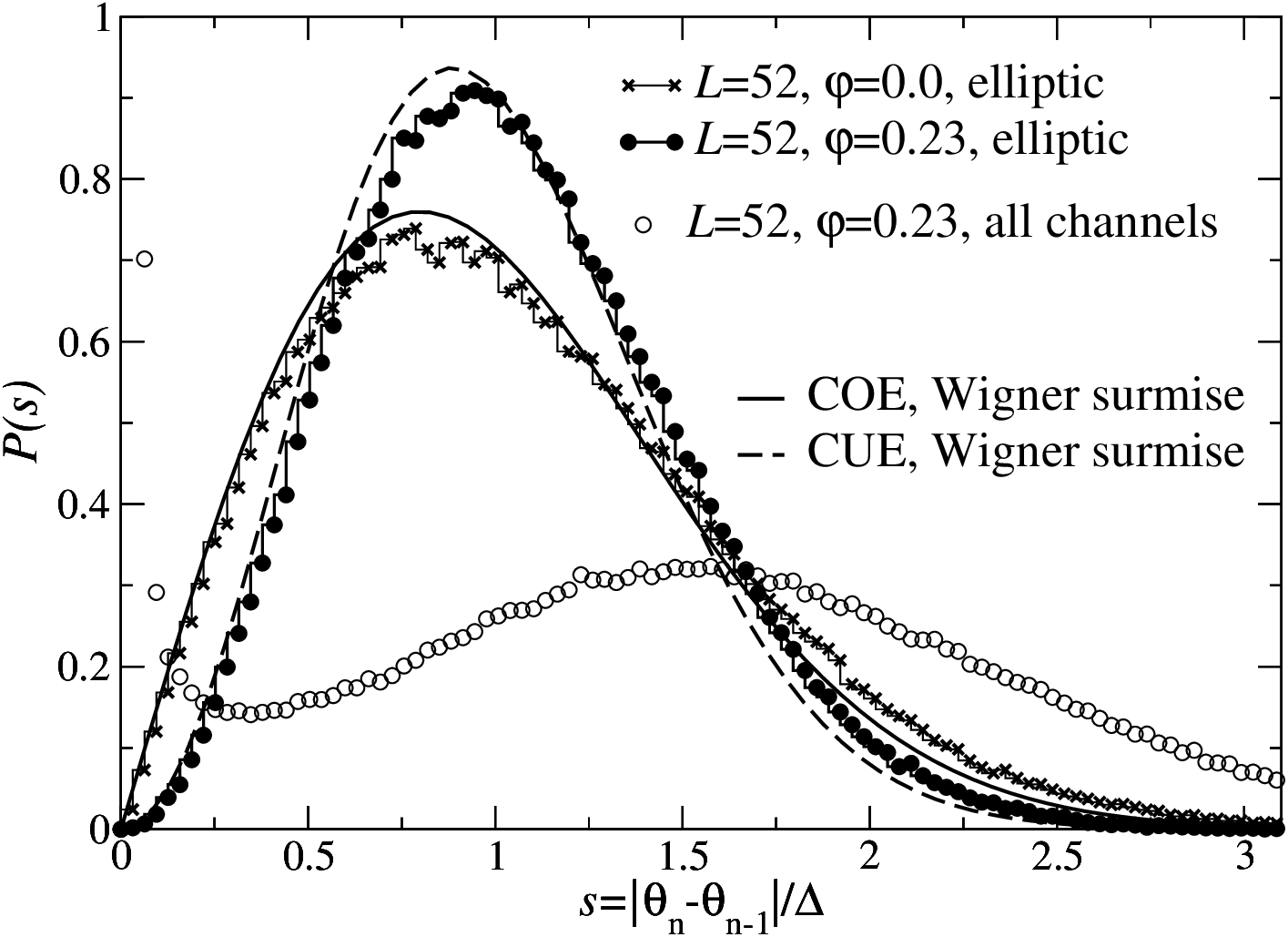}
\caption{\label{fig-5}{
Plot of the level spacing $P(s)$ of $\pi_e U_N(V_N)^*\pi_e$ generated for the Anderson model at $\varphi=0$ ($\times$) and $\varphi/2\pi=0.23$ ($\bullet$), $\lambda=0.12/\sqrt{12}$ and $E=1.31$ on a slab of width $L=52$ and length $N=1000$. The small open circles ($\circ$) show $U_N(V_N)^*$, i.e.\ without removing any hyperbolic channels. The lines denote the Wigner surmises.}}
\end{center}
\end{figure}

As $S_2+S_2^*$ is real symmetric, it can be diagonalized by an orthogonal (unitary with real entries) which we construct first.  Note that the fundamental (in our choice of sign of $S_2+S_2^*$ lying on top of the spectrum) is always non-degenerate and has always a real eigenvector $\widehat{f}_L=f_L$; moreover, for even $L$, the eigenvector $\widehat{f}_{L/2}=f_{L/2}$ is real as well and has a non-degenerate eigenvalue. For other $l<\frac{L}{2}$, real normalized eigenvectors $\widehat{f}_l$ are obtained by
\begin{equation}
\label{eq-realFourier}
(\widehat{f}_l,\widehat{f}_{L-l}) \;=\; (f_l,f_{L-l})\;\frac{1}{\sqrt{2}}\, \left(
\begin{array}{cc}
-\imath & 1 \\
\imath & 1
\end{array}
\right) \mbox{ . }
\end{equation}
\noindent This is just the Fourier basis given by cosines and sinus. Now set
$$
\widehat{m}\;=\;(\widehat{f}_1,\ldots,\widehat{f}_L)\in\mbox{O}(L)\mbox{ . }
$$
There exists a unitary matrix $a\in\,$U$(L)$ essentially given by the r.h.s. of \eqref{eq-realFourier} such that $\widehat{m}=ma$
where $m$ is the discrete Fourier transform as defined in Section~\ref{sec-normal}. 
%
%
%
%
If one now associates to $\widehat{m}$ as before a matrix $\widehat{\Mm}=\left(\begin{smallmatrix} \widehat{m} & 0 \\ 0 & \widehat{m} \end{smallmatrix}\right)$ and carries out the associated basis change as in \eqref{eq-transfer2}, the matrices $q$ and $\mu$ remain the same (hence don't carry a sombrero) and ${W}$ is replaced by the real matrix $\widehat{W}=a^*{W}a$. As $q$ and $\mu$ are unchanged, so are the diagonal matrices $\rho$, $\pi_e$, $\pi_h$ and $g$ which all have the two-fold degeneracy (apart from the fundamentals). Also $\widehat{\Rr}=\Rr$ is unchanged, but in \eqref{eq-transfer3} the definition of $\widehat{\Pp}$ contains $\widehat{W}$ instead of $W$, namely $\widehat{\Pp}=\Aa^*\Pp\Aa$ with $\Aa=\,$diag$(a,a)$.  With these changes, one has

$$
\widehat{\Tt}\;=\;
\widehat{\Rr}\,e^{\lambda\widehat{\Pp}}\;\in\; \mbox{HS}(2L,\RM)\;,
\qquad \widehat{\Pp}\in \mbox{hs}(2L,\RM)\;.
$$
Again the RPP can be checked numerically in a similar fashion as in Section~\ref{sec-normal}. The results are of the same nature. In order to exhibit the difference between the cases with and without magnetic field, we plot the distribution of the level spacing of $\pi_e U_N\pi_e(\pi_e V_N\pi_e)^*\approx \pi_e U_N(V_N)^*\pi_e$ for the case with and without magnetic field. For $\varphi\not = 0$ one has CUE statistics reflecting that $U_N$ and $V_N$ are essentially independent, while for $\varphi=0$ the relation $V_N=U_N$ implies COE statistics, both facts that can be read off Figure~\ref{fig-5}.

\vspace{.1cm}

Now the calculation of the Lyapunov spectrum using case $\KM=\RM$ of Theorem~\ref{theo-equidistance}, we need to calculate $\EE\,\Tr(\Pi_e(\widehat{\Pp}^*+\widehat{\Pp})\,\Pi_e\,\widehat{\Pp}\,\Pi_e)$.
But the unitary $\Aa$ commutes with $\Pi_e$ so that the result is given by the formula \eqref{eq-expval}. From this follows formula \eqref{eq-gammapreal}.

\section{Ando model on tube geometry}
\label{sec-Ando}

In this section, the symplectic universality class is dealt with. Hence we take $H=H_0+\lambda H_1$ with $\varphi=0$ and $t\not =0$ which is the Ando Hamiltonian \cite{Ando}. Of course, the time reversal invariance (TRI) of $H$ is a crucial property. Recall that for systems with odd spin (actually half odd spin) the time reversal operator is given by complex conjugation followed by a rotation in spin space by 180 degrees, which is given by the operator $I=e^{\imath\pi s^y}=\left(\begin{smallmatrix} 0 & -1 \\ 1 & 0 \end{smallmatrix}\right)$, just as it was defined in Section~\ref{sec-action}. This operator acts on the spin variable only and could hence also be written as $\one\otimes I$ in $\ell^2(\ZM,\CM^L)\otimes \CM^2$. The TRI property of $H$ now reads $I^*\overline{H}I=H$ which can readily be verified using the identity $I^*\overline{{\bf s}}I=-{\bf s}$. As recalled in Section~\ref{sec-action}, this means precisely that $H$ has quaternion entries. This just reflects that the spin operators are linked to the quaternions by $(q_3,q_2,q_1)=2\imath {\bf s}$.
Similar as the objects without magnetic field carried a sombrero, we place here a tilde on the objects linked to the Ando model.

\vspace{.1cm}

The transfer matrix $\Ss$ associated to the Ando Hamiltonian is given in \eqref{eq-transferintro}.
For the analysis of the normal form, it is convenient to think of it as a $2L\times 2L$ matrix with real quaternion entries. In fact, one may write
\begin{equation}
\label{eq-transferAndo2} \widetilde{\Ss} \;=\; \left(
\begin{array}{cc} [E -(S_2+S_2^*)-t(S_2-S_2^*)q_3](1+tq_2)^{-1} & \! -(1+tq_2)^* \\
(1+tq_2)^{-1} &\!  {0}\end{array} \right)
\left(
\begin{array}{cc} \one & \! 0 \\
\lambda (1+t^2)^{-1}w & \! \one\end{array} \right)
\,,
\end{equation}
and the TRI is then simply the fact that $\widetilde{\Ss}\in\mbox{HS}(2L,\HM)$ can indeed be written using the quaternions with real coefficients.

\vspace{.1cm}

Next we want to bring $\widetilde{\Ss}$ for $\lambda=0$ into a normal form similar as in Section~\ref{sec-normal}. A first step can be to apply the real Fourier transfer $\widehat{\Mm}$ already used in Section~\ref{sec-zeromag}. As it is real, this conserves the TRI. The result is the analog of \eqref{eq-transfer2}:
\begin{equation}
\label{eq-transferAndo3}
\widehat{\cal M}^{-1}\widetilde{\Ss}\widehat{\cal M}
\;=\; \left(
\begin{array}{cc}  \widetilde{\mu}_0 (1+tq_2)^{-1}& -(1+tq_2)^* \\
(1+tq_2)^{-1} & 0\end{array} \right) \;\left(
\begin{array}{cc}  {\bf 1} & 0 \\
\lambda(1+t^2)^{-1}\,\widehat{W} & {\bf 1}\end{array} \right)
\;,
\end{equation}
where $\widehat{W}$ is as in Section~\ref{sec-zeromag} and $\widetilde{\mu}_0
=\widehat{m}^*(E -(S_2+S_2^*)-t(S_2-S_2^*)q_3)\widehat{m}$. Due to conserved TRI, all matrices appearing in \eqref{eq-transferAndo3} are in $\mbox{HS}(2L,\HM)$. However, $\widetilde{\mu}_0$ is not diagonal any more as it was the case without spin-orbit coupling. Indeed, it is a direct sum of (the fundamentals and) $2\times 2$ quaternion blocks of the form
$$
\left(
\begin{array}{cc}  E-2\cos(\widetilde{\eta})& -2t\sin(\widetilde{\eta})\,q_3 \\
2t\sin(\widetilde{\eta})\,q_3 & E-2\cos(\widetilde{\eta})\end{array} \right)
\;,
$$
with a frequency $\widetilde{\eta}\in \frac{2\pi}{L}\,\ZM$ which can be calculated for each block. In \eqref{eq-transferAndo3} this leads to $4\times 4$ quaternion blocks in $\mbox{HS}(4,\HM)$, which are thus $8\times 8$ matrices with complex entries. It is clear that diagonalizing them is a hard task. It is somewhat easier to first apply only the complex Fourier transform $\Mm=\widehat{\Mm}\Aa^*$:
\begin{eqnarray}
\label{eq-transferAndo4}
{\cal M}^{-1}\widetilde{\Ss}{\cal M}
&=& \left(
\begin{array}{cc} [E-2\cos(\eta)+2\imath t \sin(\eta)q_3]  (1+tq_2)^{-1}& \!-(1+tq_2)^* \\
(1+tq_2)^{-1} & \!0\end{array} \right) \times \nonumber \\
& & \mbox{ } \left(
\begin{array}{cc}  {\bf 1} & \! 0 \\
\lambda(1+t^2)^{-1}\,{W} & \!{\bf 1}\end{array} \right)
.
\end{eqnarray}
Indeed, now the upper left entry and $\widetilde{\mu}=E-2\cos(\eta)+2\imath t \sin(\eta)q_3$ have only $2\times 2$ blocks on the diagonal. These diagonals are not quaternions though (due to the imaginary coefficients of $q_3$) and, in fact, the matrices on the r.h.s. are only in $\mbox{HS}(4L,\CM)$ and not in $\mbox{HS}(2L,\HM)$. On the other hand, the lowest order in $\lambda$ now has $4\times 4$ complex blocks and is thus easier to diagonalize. It is then possible to recombine these diagonalized blocks to reinstall TRI. This diagonalization is carried out in detail in Appendix B. The result is the construction of a matrix $\widetilde{\Nn}\in \mbox{HS}(2L,\HM)$ such that
$$
\widetilde{\Nn}^{-1}\widetilde{\Qq}^{-1}\widehat{\Mm}^{-1}
\widetilde{\Ss}\widehat{\Mm}\widetilde{\Qq}\widetilde{\Nn}
\;=\;
\widetilde{\Rr}\,e^{\lambda\widetilde{\Pp}}
\;,
$$
where now all factors are in $\mbox{HS}(2L,\HM)$ and $\widetilde{\Rr}$ is in a normal form. Therefore, one is precisely in the situation where the RPP for the case $\KM=\HM$ may hold and the Lyapunov spectrum can be calculated by Theorem~\ref{theo-equidistance}. In order to write out a formula for the Lyapunov exponents, one needs to calculate $\EE\,\Tr(\Pi_e(\widetilde{\Pp}^*+\widetilde{\Pp})\,\Pi_e\,\widetilde{\Pp}\,\Pi_e)$. Because of the complicated form of $\widetilde{\Nn}$ (given in Appendix B) this is quite a formidable algebraic challenge, which we refrain from dealing with for the following reason. The matrix $\widetilde{\Pp}$ and hence the coefficient in
Theorem~\ref{theo-equidistance} depends in a differentiable way on $t$, hence for small $t$ the coefficient is given by the one in the case $\KM=\RM$. Therefore the formula \eqref{eq-gammapquat} follows.

\section{Anderson model on a slab}
\label{sec-slab}

As an approximation to a $d$-dimensional situation, let now consider operators on the Hilbert space $\ell^2(\ZM)\otimes\CM^N)^{d-1}$. Here one may view $(\CM^N)^{d-1}$ as the Hilbert space of a $(d-1)$-dimensional discrete torus and the total physical space $\ZM\times\{1,\ldots,N\}^{d-1}$ as a discretized and periodized slab. The dimension of the fiber is here $L=N^{d-1}$. On the fiber there are cyclic shifts $S_2,\ldots,S_d$ satisfying each $S_j^N=\one$. Given magnetic phases $\varphi_2,\ldots,\varphi_d$, the kinetic operator is now

$$
H_0\;=\;S_1+S_1^*+\Delta\;,
\qquad
\Delta
\;=\;
\sum_{j=2}^d\bigl(e^{\imath\varphi_j}S_j+e^{-\imath\varphi_j}S_j^*\bigr)\;.
$$
The diagonal disorder is $V=\lambda\sum_{n_1\in\ZM}w_{n_1}$ with $w_{n_1}= \sum_{n_2,\ldots, n_d=1}^Nw_n\,|n\rangle\langle n|$ where the $w_n$'s are i.i.d. real and centered random variables with unit variance and $n=(n_1,\ldots,n_d)$. Viewing $\Delta$ also as an operator on $\CM^L$, the transfer matrix now is $\Ss=
\left(\begin{array}{cc}E-\Delta-\lambda w_{n_1} & -\one \\ \one & 0 \end{array}\right)$. Its normal form is found exactly in the same way as in Section~\ref{sec-normal}. Let $m_j$ for $j=2,\ldots,d$ be the discrete Fourier transforms in the transverse directions and set $m=m_2\otimes\ldots\otimes m_d$. Then $m$ is an $L\times L$ unitary matrix diagonalizing $\Delta$. Thus also $m^*(E-\Delta)m$ is diagonal. The ordering
according to the modulus is again obtained by some permutation $q$. Then set $\mu=q^*m^*(E-\Delta)mq$.  Then define $\pi_e$, $\pi_h$, $\rho$, $h=h(\varphi)$, $\Nn$ and $\Pi_e$ using the same formulas as in Section~\ref{sec-normal}. This leads again to a normal form with a new perturbation $\Pp$. One can now carefully check that \eqref{eq-expval} holds again, with $h$ as just defined and $L=N^{d-1}$. Therefore Theorem~\ref{theo-equidistance} shows (say in the case of the unitary universality class) that

$$
\gamma^\CM_p
\;=\;
\frac{\lambda^2}{4N^{d-1}}\;
\left(\frac{1}{L_e}\sum_{k=L_h+1}^L
h_k^2\right)^2\;\left(L-p+\frac{1}{2}\right)\;+\;\Oo(\lambda^3)\;.
$$
Note that the number of terms in the sum over $k$ is exactly $L_e$ and each summand is of order $1$, so that the inverse localization length is roughly equal to $\gamma^\CM_L\sim\frac{\lambda^2}{4N^{d-1}}$.


\section*{Appendix A: Moments of the Haar measure on $\mbox{\rm U}(L)$}

In this appendix, we reassemble some results, most well-known, about second and fourth moments of the Haar measure on the unitary group $\mbox{\rm U}(L)$. The notation for the entries of a unitary matrix is $U=(U_{p,q})_{1\leq p,q\leq L}$ and the average w.r.t. to the Haar measure is denote by $\langle\,.\,\rangle$. We will need formulas for the low moments of unitaries. All first and third moments vanish, and the only non-vanishing second and fourth moments are given by the following list (given, {\it e.g.}, in \cite{HP}):

\vspace{.2cm}

\noindent (M1) $\langle \overline{U_{p,q}}U_{k,l}\rangle=\frac{1}{L}$ for $p=k$ and $q=l$.

\vspace{.1cm}

\noindent (M2) $\langle \overline{U_{k,p}}\overline{U_{l,q}}U_{m,q}U_{n,p}\rangle =\frac{1}{L^2-1}$ for $k=n\neq l=m$ and $p\neq q$.

\vspace{.1cm}

\noindent (M3) $\langle \overline{U_{k,p}}\overline{U_{l,q}}U_{m,q}U_{n,p}\rangle =-\frac{1}{L(L^2-1)}$ for $k=m\neq l=n$ and $p\neq q$.

\vspace{.1cm}

\noindent (M4) $\langle \overline{U_{k,p}}\overline{U_{l,q}}U_{m,q}U_{n,p}\rangle =\frac{1}{L(L+1)}$ for $k=n= l=m$ and $p\neq q$.

\vspace{.1cm}

\noindent (M5) $\langle \overline{U_{k,p}}\overline{U_{l,q}}U_{m,q}U_{n,p}\rangle =\frac{1}{L(L+1)}$ for either $k=n\neq l=m$ or $k=m\neq l=n$, both with $p=q$.

\vspace{.1cm}

\noindent (M6) $\langle \overline{U_{k,p}}\overline{U_{l,q}}U_{m,q}U_{n,p}\rangle =\frac{2}{L(L+1)}$ for $k=n= l=m$ and $p=q$.

\vspace{.2cm}

For our purposes, it will be convenient to deduce the following identities.

\begin{lemma}
\label{lem-moments} Let $A,B,C,D\in\mbox{\rm Mat}(L\times L,\CM)$. Then
\begin{eqnarray*}
\mbox{\rm (i) }\;\;\;\;\;\;\;\;\;\;\;\;\;\;\;\;
\langle\,\Tr(U^*AUB)\,\rangle
& = &
\frac{1}{L}\,\Tr(A)\Tr(B)
\\
\mbox{\rm (ii) }\;\;\;\;\;\;\;\;\;\;\;\;\;\;\;\;\;
\langle\,\Tr(\overline{U}AUB)\,\rangle
& = &
\frac{1}{L}\,\Tr(AB^t)
\\
\mbox{\rm (iii) }\;\,
\langle\,\Tr(U^*AUBU^*CUD)\,\rangle
& = &
\frac{1}{L^2-1}\Big[\Tr(A)\Tr(C)\Tr(BD)+\Tr(AC)\Tr(B)\Tr(D)\Big]
\\
& & -\;
\frac{1}{L(L^2-1)}\Big[\Tr(AC)\Tr(BD)+\Tr(A)\Tr(B)\Tr(C)\Tr(D)\Big]
\;,
\\
%
\mbox{\rm (iv) }\;\;
\langle\,\Tr(U^*AUBU^tC\overline{U}D)\,\rangle
& = &
\frac{1}{L^2-1}\Big[\Tr(A)\Tr(C)\Tr(BD)+\Tr(AC^t)\Tr(BD^t)\Big]
\\
& & -\;
\frac{1}{L(L^2-1)}\Big[\Tr(AC^t)\Tr(BD)+\Tr(A)\Tr(C)\Tr(BD^t)\Big]
\;,
\\
\mbox{\rm (v) }\;\;
\langle\,\Tr(U^*A\overline{U}BU^tCUD)\,\rangle
& = &
\frac{1}{L^2-1}\Big[\Tr(AC^t)\Tr(BD^t)+\Tr(AC)\Tr(B)\Tr(D)\Big]
\\
& & -\;
\frac{1}{L(L^2-1)}\Big[\Tr(AC)\Tr(BD^t)+\Tr(AC^t)\Tr(B)\Tr(D)\Big]
\;.
\end{eqnarray*}
Moreover, all formulas remain valid if all $U$'s are replaced by their complex conjugates, {\it e.g.}
$$
\langle\,\Tr(U^tA\overline{U}BU^tC\overline{U}D)\,\rangle
\;=\;
\langle\,\Tr(U^*AUBU^*C{U}D)\,\rangle
\;.
$$

\end{lemma}

\noindent {\bf Proof.} Item (i) and (ii) follow directly from (M1). Let us exemplify the proof of the others by treating (iii) using (M2)-(M6). First write out the trace explicitly:
$$
\Tr(U^*AUBU^*C{U}D) \;= \;
\sum_{p,q,p',q'}\sum_{k,l,m,n}
\overline{U_{k,p}}A_{k,l}U_{l,p'}B_{p',q}\overline{U_{m,q}}C_{m,n}U_{n,q'}D_{q',p}
\;,
$$
where all sums run from $1$ to $L$.
Now for a non-vanishing average one needs either $(p',q')=(p,q)$ or $(p',q')=(q,p)$. Hence
\begin{eqnarray*}
\langle\Tr(U^*AUBU^*C{U}D)\rangle & = &
\sum_{p\not =q}\sum_{k,l,m,n}
\langle\overline{U_{k,p}}U_{l,p}\overline{U_{m,q}}U_{n,q}\rangle
\;A_{k,l}B_{p,q}C_{m,n}D_{q,p}
\\
& & +\;
\sum_{p\not =q}\sum_{k,l,m,n}
\langle\overline{U_{k,p}}U_{l,q}\overline{U_{m,q}}U_{n,p}\rangle
\;A_{k,l}B_{q,q}C_{m,n}D_{p,p}
\\
& & +\;
\sum_{p}\sum_{k,l,m,n}
\langle\overline{U_{k,p}}U_{l,p}\overline{U_{m,p}}U_{n,p}\rangle
\;A_{k,l}B_{p,p}C_{m,n}D_{p,p}
\;.
\end{eqnarray*}
Let us call the three summands on the r.h.s. $I_1$, $I_2$ and $I_3$.
Next using (M2), (M3) and (M4) and the identity $\frac{1}{L^2-1}-\frac{1}{L(L^2-1)}=\frac{1}{L(L+1)}$, one has
$$
I_1\;=\;
\frac{1}{L^2-1}
\sum_{p\not =q}\sum_{k,m}
\;A_{k,k}B_{p,q}C_{m,m}D_{q,p}
\;-\;\frac{1}{L(L^2-1)}
\sum_{p\not =q}\sum_{k,m}
\;A_{k,m}B_{p,q}C_{m,k}D_{q,p}
\;.
$$
Apart from the missing diagonal term $p=q$, this gives the summands
$\frac{1}{L^2-1}$ $\Tr(A)$ $\Tr(C)$ $\Tr(BD)$ and $\frac{-1}{L(L^2-1)}\Tr(AC)\Tr(BD)$ on the r.h.s. of (iii). Similarly, the two other summands of (iii) are given by $I_2$ again up to the missing diagonal terms. However, $I_3$ can be evaluated using (M5) and (M6). Due to the two cases in (M5) and the factor $2$ in (M6), the contribution of $I_3$ provides precisely both of the diagonal terms, as can be checked using once again the identity of fractions cited above.
\hfill $\Box$

\section*{Appendix B: Transfer matrix of spin orbit Laplacian}

In this appendix we diagonalize
$$
\Ss_\eta
\;=\;
\left(
\begin{array}{cc} (E-2\cos(\eta)+2\imath t\sin(\eta)q_3)  (1+tq_2)^{-1}& -(1+tq_2)^* \\
(1+tq_2)^{-1} & 0\end{array} \right)
\;\in\;
\mbox{\rm HS}(4,\CM)
\;.
$$
This is one of the $4\times 4$ blocks of \eqref{eq-transferAndo4}. Hence we may think of $\eta$ as a real number, even though it will be a diagonal matrix in Section~\ref{sec-Ando}. Note that even though $\Ss_\eta$ is written with quaternions representing $2\times 2$ blocks, it is not a quaternion matrix because there is an imaginary coefficient. Another important fact is that $\Ss_\eta$ and $\Ss_{-\eta}$ are similar, namely
\begin{equation}
\label{eq-similarity}
\left(
\begin{array}{cc} q_2 & 0 \\
0 & q_2 \end{array} \right)^{-1}
\,\Ss_{-\eta}\,
\left(
\begin{array}{cc} q_2 & 0 \\
0 & q_2 \end{array} \right)
\;=\;
\Ss_\eta
\;.
\end{equation}
Therefore, the spectra of $\Ss_\eta$ and $\Ss_{-\eta}$ are equal and their eigenvectors linked by a simple relation. This then allows to build a matrix in $\mbox{\rm HS}(4,\HM)$ from the checker board direct sum of $\Ss_\eta$ and $\Ss_{-\eta}$.

\vspace{.1cm}

Setting $e=E-2\cos(\eta)$ and $f=2t\sin(\eta)$, the matrix $\Ss_\eta $ can be written out in a more explicit way
\begin{equation}
\label{eq-transferx}
\Ss_\eta\;=\;
\begin{pmatrix}
\frac{e-ft}{1+t^2} & \frac{-et-f}{1+t^2} & -1 & t \\
\frac{et-f}{1+t^2} & \frac{e+ft}{1+t^2} & -t & -1 \\
\frac{1}{1+t^2} & \frac{-t}{1+t^2} & 0 & 0 \\
\frac{t}{1+t^2} & \frac{1}{1+t^2} & 0 & 0
\end{pmatrix}
\;.
\end{equation}
The characteristic polynomial is real and given by
$$
p(\lambda)=\lambda^4-a\lambda^3+b\lambda^2-a\lambda+1
\;=\;
\lambda^2\left[
(\lambda+\lambda^{-1})^2-a(\lambda+\lambda^{-1})+b-2
\right]
\;,
$$
where
$$
a
\;=\;
\frac{2\,e}{1+t^2}\;,
\qquad
b
\;=\;
\frac{e^2t^2-f^2+2-2\,t^4}{(1+t^2)^2}
\;.
$$
Characteristic polynomials of this form always appears for real symplectic matrices and the stability analysis is carried out in a standard manner (as described,{\it e.g.}, in \cite{HM}). Solving the quadratic equation in $\lambda+\lambda^{-1}$ first shows
$$
p(\lambda)
\;=\;
\bigl(\lambda^2-\nu_+\lambda+1\bigr)
\bigl(\lambda^2-\nu_-\lambda+1\bigr)
\;,
\qquad
\nu_\pm
\;=\;
\frac{a}{2}\pm\sqrt{\frac{a^2}{4}+2-b}
\;.
$$
Therefore the four eigenvalues of $\Ss_\eta$ are $\kappa_+,\frac{1}{\kappa_+},\kappa_-,\frac{1}{\kappa_-}$ where
$$
\kappa_\pm
\;=\;
\frac{\nu_\pm}{2}\,+\,\sqrt{\frac{\nu_\pm^2}{4}-1}
\;.
$$
In these formulas the square root is taken to be the first branch so that both $\kappa_\pm$ have non-negative imaginary part.
Except for a discrete set of energies, all four eigenvalues are distinct. We restrict our attention to this case.
One may then think of $\Ss_\eta$ as being composed of two channels given by the two eigenvalue pairs $\kappa_+,\frac{1}{\kappa_+}$ and $\kappa_-,\frac{1}{\kappa_-}$ and the corresponding two-dimensional eigenspaces. If $\kappa_\pm\in\SM^1$, the channel is elliptic or parabolic, otherwise hyperbolic. Whenever $\frac{a^2}{4}+2-b<0$, one has $\nu_+=\overline{\nu_-}$ and both channels are hyperbolic but, in fact, not uncorrelated because $\kappa_+=(\overline{\kappa_-})^{-1}$. Whenever $\frac{a^2}{4}+2-b>0$, $\kappa_\pm$  is elliptic if and only if $|\nu_\pm|<2$. Resuming, the eigenvalues are generically in one of the following four constellations:

\vspace{.1cm}

\noindent (G1) $\kappa_+=(\overline{\kappa_-})^{-1}$ with $\kappa_+\not\in \SM^1\cup\RM$.

\vspace{.1cm}

\noindent (G2) $\kappa_\pm\in\SM^1/\{\pm 1\}$ with $\kappa_+\not =\kappa_-$.

\vspace{.1cm}

\noindent (G3) $\kappa_+\in\SM^1/\{\pm 1\}$ and $\kappa_-\in\RM/\{\pm 1\}$ or $\kappa_-\in\SM^1/\{\pm 1\}$ and $\kappa_+\in\RM/\{\pm 1\}$.

\vspace{.1cm}

\noindent (G4)  $\kappa_\pm\in\RM/\{\pm 1\}$.

\vspace{.1cm}

\noindent The cases (G1) and (G4) are fully hyperbolic, in (G3) there is an elliptic and a hyperbolic channel, while (G2) is fully elliptic.

\vspace{.2cm}

Next let us determine eigenvectors $v_{+},v'_{+},v_{-},v'_{-}\in\CM^4$  associated in that order to the eigenvalues $\kappa_+,\frac{1}{\kappa_+},\kappa_-,\frac{1}{\kappa_-}$. Due to the simple form \eqref{eq-transferx} of $\Ss_\eta$, the lowest two components of the eigenvalue equation allow to express the last two components of the eigenvectors in terms of the first two. Replacing this, the first two components of the eigenvalue equation become equations in the first two components of eigenvalues only. One of them is redundant because of the eigenvalue property. With some care, one thus gets
$$
v_{\pm}
\;=\;
c_\pm\,\begin{pmatrix}
2\kappa_{\pm}t-\kappa_{\pm}^2(et+f)
\\
-2\kappa_{\pm}t^2+ \kappa_{\pm}^2t(et+f)
\\
2t-\kappa_{\pm}f-\kappa_{\pm}^2t
\\
-\kappa_{\pm}e +\kappa_{\pm}^2
\end{pmatrix}
\;,
$$
where $c_\pm\in\CM$ are normalization constants to be chosen later. Using $\frac{1}{\kappa_{\pm}}$ instead of $\kappa_{\pm}$, the eigenvectors $v'_\pm$ are constructed similarly with normalization constants $c'_\pm$. As all of these are eigenvectors of a hermitian symplectic matrix, they have isotropy properties which we exemplify next (an exhaustive list is then readily made). One has $(v_\pm)^*\Jj v_\pm=|\kappa_\pm|^{-2}(\Ss_\eta v_\pm)^*\Jj \Ss_\eta v_\pm
=|\kappa_\pm|^{-2}(v_\pm)^*\Jj v_\pm$. Hence whenever $\kappa_\pm\not\in\SM^1$ one has $(v_\pm)^*\Jj v_\pm=0$. Similarly $(v_\pm)^*\Jj v_\mp=0$ whenever $\overline{\kappa_\pm}\kappa_\mp\not = 1$, $(v_\pm)^*\Jj v'_\pm=0$ whenever $\overline{\kappa_\pm}(\kappa_\pm)^{-1}\not = 1$ and
$(v_\pm)^*\Jj v'_\mp=0$ whenever $\overline{\kappa_\pm}(\kappa_\mp)^{-1}\not = 1$. Note that this gives no information on $(v_\pm)^*\Jj v_\pm$ when $\kappa_\pm\in\SM^1$. However, the antisymmetry of $\Jj$ always guarantees that
$(v_\pm)^*\Jj v_\pm\in\imath\,\RM$. Another important fact resulting from the reality of $\Ss_\eta$ is that if $\Ss_\eta v_\pm=\kappa_\pm v_\pm$, then also
$\Ss_\eta \overline{v_\pm}=\overline{\kappa_\pm}\, \overline{v_\pm}$. This may be empty if $v_\pm$ is real (which can only happen for real $\kappa_\pm$), but produces a relation between eigenvectors otherwise.

\vspace{.1cm}

From the eigenvectors we now build a matrix $\Mm_\eta\in\mbox{\rm HS}(4,\CM)$ such that $\Mm_\eta^{-1}\Ss_\eta\Mm_\eta$ is in a normal form. This has to be done in each of the case (G1) to (G4) separately.

\vspace{.1cm}

\noindent (G1) Here $\overline{v_\pm}=v'_\mp$. Furthermore the pairs $(v_+,\overline{v_+})$ and $(v_-,\overline{v_-})$ span a Lagrangian plane (maximally hermitian isotropic for $\Jj$). The normalization constants can be chosen such that $(v_+)^*\Jj v_-=1$. Then set $\Mm_\eta=(v_+,{v'_-},v_-,{v'_+})=(v_+,\overline{v_+},v_-,\overline{v_-})$. One can check that $\Mm_\eta\in\mbox{\rm HS}(4,\CM)$ and $(\Mm_\eta)^{-1}\Ss_\eta \Mm_\eta=\mbox{\rm diag}(\kappa_+,\overline{\kappa_+},(\kappa_+)^{-1},(\overline{\kappa_+})^{-1}) \in\mbox{\rm HS}(2,\HM)$. Below we will use that also the complex conjugate
$\overline{\Mm_\eta}\in\mbox{\rm HS}(4,\CM)$ diagonalizes $S_\eta$.

\vspace{.1cm}

\noindent (G2) Here the eigenvalues are $e^{\imath\varphi_+},e^{-\imath\varphi_+},e^{\imath\varphi_-},e^{-\imath\varphi_-}$ with eigenvectors $v_+,\overline{v_+},v_-,\overline{v_-}$. Neither of these eigenvectors are isotropic. Thus we introduce the real vectors $w_\pm=\Re e(v_\pm)$ and $w'_\pm=\Im m(v_\pm)$. Hence all the vectors $w_\pm,w'_\pm$ are isotropic (because $(w_\pm)^*\Jj w_\pm$ is both real and imaginary). Furthermore $(w_\pm)^*\Jj w_\mp=0$ and $(w_\pm)^*\Jj w'_\mp=0$. By adequate choice of the normalization constants one can achieve $(w_\pm)^*\Jj w'_\pm=0$. All this assures that $\Mm_\eta=(w_+,w_-,w'_+,w'_-)\in\mbox{\rm HS}(4,\RM)$. Furthermore, one can check
\begin{equation}
\label{eq-G2diag}
(\Mm_\eta)^{-1}\Ss_\eta \Mm_\eta
\;=\;
\begin{pmatrix}
\cos(\varphi_+) & 0 & - \sin(\varphi_+) & 0 \\
0 & \cos(\varphi_-) & 0 & - \sin(\varphi_-)  \\
\sin(\varphi_+) & 0 & \cos(\varphi_+) & 0 \\
0 & \sin(\varphi_-) & 0 & \cos(\varphi_-)
\end{pmatrix}\;
\in\;\mbox{\rm HS}(4,\CM)
\;.
\end{equation}

\vspace{.1cm}

\noindent (G3) This is a combination of (G2) above and (G4) treated next.

\vspace{.1cm}

\noindent (G4) Here all eigenvectors are real and thus isotropic, and, moreover,
one has $(v_\pm)^*\Jj v'_\mp=0$, $(v_\pm)^*\Jj v_\mp=0$ and $(v'_\pm)^*\Jj v'_\mp=0$. Hence $(v_+)^*\Jj v_-$ and $(v'_+)^*\Jj v'_-$ cannot vanish (otherwise there would be a three-dimensional isotropic subspace) and can by choice of the normalization constants both be made equal to $1$. Then $\Mm_\eta=(v_+,v'_-,v'_+,v_-)\in\mbox{\rm HS}(4,\RM)$ and one can verify $(\Mm_\eta)^{-1}\Ss_\eta \Mm_\eta=\mbox{\rm diag}(\kappa_+,(\kappa_-)^{-1},(\kappa_+)^{-1},\kappa_-) \in\mbox{\rm HS}(2,\CM)$.
Note that the complex conjugate of the latter diagonal matrix is also in $\mbox{\rm HS}(2,\CM)$.

\vspace{.2cm}

In each of the four cases let us denote $\Dd_\eta\in\mbox{\rm HS}(4,\CM)$ the matrix such that
$$
(\Mm_\eta)^{-1}\Ss_\eta \Mm_\eta
\;=\;
\Dd_\eta
\;,
\qquad
(\overline{\Mm_\eta})^{-1}\Ss_\eta \overline{\Mm_\eta}
\;=\;
\overline{\Dd_\eta}
\;.
$$
Note that the complex conjugate equation differs from the first one only in case (G1). Further by \eqref{eq-similarity},
$$
(\Mm_{-\eta})^{-1}\Ss_{-\eta} \Mm_{-\eta}
\;=\;
\Dd_\eta
\;,
\qquad
\Mm_{-\eta}
\;=\;
\begin{pmatrix}
q_2 & 0 \\
0 & q_2
\end{pmatrix}
\,\Mm_{\eta}
\;.
$$
We apply the checker board sypmplectic sum  
to the matrices $\Ss_\eta$ and $\Ss_{-\eta}$ with quaternion entries (but with complex coefficients), and consider also the $4\times 4$ matrix $\Mm_\eta$ as a $2\times 2$ matrix of $2\times 2$ blocks so that also $\overline{\Mm_{\eta}}\widetilde{\oplus}\Mm_{-\eta}$ is well-defined. Then the above translates to
$$
(\overline{\Mm_{\eta}}\widetilde{\oplus}\Mm_{-\eta})^{-1}
\,\Ss_{\eta}\widetilde{\oplus}\Ss_{-\eta}
\;\overline{\Mm_{\eta}}\widetilde{\oplus}\Mm_{-\eta}
\;=\;
\overline{\Dd_\eta}
\widetilde{\oplus}
\Dd_\eta
\;,
$$
and each factor is in $\mbox{\rm HS}(8,\CM)$.
Now consider similar as in Section~\ref{sec-zeromag} the matrix
$\Aa=\mbox{\rm diag}(a,a)$ with a $4\times 4$ matrix $a=2^{-\frac{1}{2}}\left(\begin{smallmatrix} -\imath & 1 \\ \imath & 1 \end{smallmatrix}\right)$, having diagonal $2\times 2$ blocks. Then $\Aa^*
\Ss_{\eta}\widetilde{\oplus}\Ss_{-\eta}\Aa\in\mbox{\rm HS}(4,\HM)$ and
$$
\bigl(\Aa^*\overline{\Mm_{\eta}}\widetilde{\oplus}\Mm_{-\eta}\Bb\bigr)^{-1}
\;
\bigl(\Aa^*\Ss_{\eta}\widetilde{\oplus}\Ss_{-\eta}\Aa\bigr)
\;\bigl(\Aa^*\overline{\Mm_{\eta}}\widetilde{\oplus}\Mm_{-\eta}\Bb\bigr)
\;=\;
\bigl(\Bb^{-1}\overline{\Dd_\eta}
\widetilde{\oplus}
\Dd_\eta\Bb\bigr)
\;,
$$
with $\Bb\in\mbox{\rm HS}(8,\CM)$ chosen such that all four factors are in $\mbox{\rm HS}(4,\HM)$:
$$
\Bb\;=\;\mbox{\rm diag}(b,b)\;,
\qquad
b\;=\;\imath\;
\begin{pmatrix}
1 & 0 & 0 & 0 \\
0 & 0 & 1 & 0 \\
0 & 1 & 0 & 0 \\
0 & 0 & 0 & 1
\end{pmatrix}
\;.
$$
In case (G2), $\Bb^{-1}\Dd_\eta
\widetilde{\oplus}
\Dd_\eta\Bb$ is given by the r.h.s. of \eqref{eq-G2diag}, albeit each entry multiplied by $q_0$. The cases (G3) and (G4) are similar, but in case (G1) one needs to use that already $\Dd_\eta\in\mbox{\rm HS}(2,\HM)$. As to $\Aa^*\overline{\Mm_{\eta}}\widetilde{\oplus}\Mm_{-\eta}\Bb$, it is in $\mbox{\rm HS}(8,\CM)$ because each factor is, and then it is tedious but straightforward to show that it actually has quaternion entries.

\vspace{.2cm}

Finally resuming, let us set $\widetilde{\Nn}_\eta=\Aa^*\overline{\Mm_{\eta}}\widetilde{\oplus}\Mm_{-\eta}\Bb$, then
$$
(\widetilde{\Nn}_\eta)^{-1}
\bigl(\Aa^*\Ss_{\eta}\widetilde{\oplus}\Ss_{-\eta}\Aa\bigr)
\;\widetilde{\Nn}_\eta
\;=\;
\Bb^{-1}\overline{\Dd_\eta}
\widetilde{\oplus}
{\Dd_\eta}\Bb
\;\in\;
\mbox{\rm HS}(4,\HM)
\;.
$$
From this one readily builds the matrix $\widetilde{\Nn}$ needed in Section~\ref{sec-Ando}.


\end{document}